\RequirePackage{fix-cm}
\documentclass[smallextended]{svjour3}       
\smartqed  
\usepackage{xcolor}
\usepackage{graphicx}
\usepackage{natbib}
\usepackage[normalem]{ulem}

\newcommand{\black}[1]{{\textcolor{black}{#1}}}
\newcommand{\add}[1]{{\black{#1}}}

\newcommand{\Ms}{M_{\rm s}}
\newcommand{\um}{\mbox{$\mu$m}}
\newcommand{\Msun}{\mbox{M$_{\odot}$}}
\newcommand{\hii}{\textsc{Hii}}

\def\msun{\ifmmode M_\odot \else M$_\odot$ \fi}
\def\rsun{\ifmmode R_\odot \else R$_\odot$ \fi}
\def\lsun{\ifmmode L_\odot \else L$_\odot$ \fi}
\def\e{\ifmmode ^{-1} \else $^{-1}$ \fi}

\def\ltsima{$\; \buildrel < \over \sim \;$}    
\def\gtsima{$\; \buildrel > \over \sim \;$}    
\def\gtrsim{\lower.5ex\hbox{\gtsima}}        
\def\lesssim{\lower.5ex\hbox{\ltsima}}      

\begin{document}

\title{Zooming in on Individual Star Formation: Low- and High-mass Stars} 



\author{Anna L. Rosen \and Stella S. R. Offner \and Sarah I. Sadavoy \and Asmita Bhandare \and Enrique V\'{a}zquez-Semadeni \and  \\ Adam Ginsburg 
}


\institute{Anna L. Rosen \at
           Center for Astrophysics $|$ Harvard \& Smithsonian \\ 
           60 Garden St., Cambridge, MA 02138, USA \\
           \email{anna.rosen@cfa.harvard.edu}           
           \and
           Stella S. R. Offner \at
              The University of Texas at Austin \\ 
              Austin, TX 78712, USA 
             \and
            Sarah I. Sadovoy \at
            Queen’s University, \\
            Kingston, ON, K7L 3N6, Canada
            \and
            Asmita Bhandare \at
            Max-Planck-Institut fur Astronomie, \\
            Konigstuhl 17, 69177 Heidelberg, Germany
            \and
            Enrique V\'{a}zquez-Semadeni \at
            Instituto de Radioastronom\'{i}a y Astrof\'{i}sica \\
            Universidad Nacional Aut\'{o}noma de M\'{e}xico \\
            Morelia, Michoac\'{a}n, 58089, M\'{e}xico
            \and
            Adam Ginsburg \at
            University of Florida \\
            Gainesville, FL 32611, USA
}

\date{Received: January 31, 2020 / Accepted: May 7, 2020}

\maketitle

\begin{abstract}
Star formation is a multi-scale, multi-physics  problem  ranging  from  the  size  scale  of  molecular clouds ($\sim10$s pc) down to the size scales of dense prestellar cores ($\sim 0.1$ pc) that are the birth sites of stars. Several physical processes like turbulence, magnetic fields and stellar feedback, such as radiation pressure and outflows, are more or less important for different stellar masses and size scales.
During the last decade a variety of technological and computing advances have transformed our understanding of star formation through the use of multi-wavelength observations, large scale observational surveys, and multi-physics multi-dimensional numerical simulations. Additionally, the use of synthetic observations of simulations have provided a useful tool to interpret observational data and evaluate the importance of various physical processes on different scales in star formation. Here, we review these recent advancements in both high- ($M \gtrsim 8 \msun$) and low-mass star formation.

\end{abstract}

\section{Introduction}
\label{intro}

Star formation is a multi-scale process that occurs in large (L $\sim$ 10 pc), dense (n $\gtrsim 10^3$ cm$^{-3}$), and cold (T $\sim$ 10 K) giant molecular clouds (GMCs) that have a predominantly hierarchical structure with increasing densities toward smaller scales that are the birth sites of stars. Stars span a large range of masses from $\sim 0.08 M_{\rm \odot}$ marking the maximum mass of brown dwarfs \add{(i.e., the minimum mass required for core deuterium burning)} to $\sim 200 M_{\rm \odot}$, the maximum stellar mass either set by stellar feedback -- the injection of energy and momentum by young stars into the interstellar medium (ISM) -- or instability \add{i.e., exploding via a pulsational pair-instability supernova once a maximum mass is reached \citep{Woosley2017a, Schneider2018a}}. Typically, high-mass and low-mass stars are separated at the mass at which stellar death results in supernovae (SNe) explosions. For simplicity, we will refer to low-mass stars as stellar products that are of insufficient mass to produce supernova events (e.g., late-type B stars or lower in mass with masses $\lesssim 8 \msun$).

Low-mass stars are the main stellar constituent of galaxies since they \add{dominate} the initial mass function (IMF) and therefore \add{dominate} the star formation process \cite[e.g.,][]{Kroupa02,Chabrier03}. They are also the sites of planet formation. In contrast, high-mass stars represent only $\sim$1\% of the stellar population in star-forming galaxies by number but they have a much more dramatic effect on their natal environments with their intense radiation fields, fast stellar winds, and subsequent SNe explosions. This feedback has direct implications for both star and galaxy formation since stellar feedback may be responsible for the dissolution of star clusters and the destruction of the GMCs out of which they form \citep{Fall2010a}. The cumulative effect of stellar feedback may also drive galactic scale outflows \citep[e.g.,][]{Geach2014a}. 


The densest condensations within GMCs are commonly referred to as prestellar cores and the gravitational collapse of these cold, dense, gaseous, and dusty cores, leads to the formation of stars. Zooming in on the smallest scales in order to understand the complex physical processes such as hydrodynamics, radiative transfer, phase transition (in particular hydrogen dissociation), and magnetic fields, entail several challenges both theoretically and observationally 
\add{\citep[e.g.][]{Nielbock2012, Launhardt2013, DunhamPPVI+2014, Wurster2018a, Teyssier2019a}}. Despite a plethora of theoretical, numerical, and observational efforts, various fundamental questions such as the values of initial magnetic field strengths and orientation, angular momenta, and turbulence of the dense cores from which stars and disks form still remain to be answered \citep[see detailed reviews by][]{Larson2003, Mckee2007,Inutsuka2012, TanPPVI+2014, Motte2018, Wurster2018a, HullZhang19, Teyssier2019a, Zhao2020}. This in turn introduces many caveats in understanding the initial conditions for star, disk, and planet formation. 

In this review we highlight the recent advances in both observations and numerical simulations in star formation for both low- and high-mass stars over the last decade. We begin with reviewing the recent advances in observations of low- and high-mass star formation in Section~\ref{sec:obs}. In Section~\ref{sec:SFtheory}, we review the theoretical and numerical studies involved with understanding both low- and high-mass star formation by focusing on the formation of individual stars. Next, we describe efforts to bridge theory and observations through ``synthetic observations," which help to interpret data, discriminate between models and evaluate the importance of various physical processes on different scales in Section~\ref{sec:synthetic}. We conclude and briefly discuss future prospects for star formation studies in Section~\ref{sec:con}.

\section{Observed Initial Conditions for Low-Mass and High-Mass Star Formation}
\label{sec:obs}

Star formation is viewed through a lens of a variety of different atoms and molecules, each with their own excitation conditions, abundances and limitations. The cold, dense conditions of prestellar (starless) cores are conducive to emission from the low-lying energy states of CO isotopologues ($^{12}$CO, $^{13}$CO, C$^{18}$O) and  nitrogen species such as NH$_3$, N$_2$H$^{+}$ and HCN. Cold conditions enhance the production of deuterated species, via the H$^+_3$ + HD $\rightarrow$ H$_2$D$^+$ + H$_2$ reaction, which ultimately leads to a relatively high abundance of molecules such as DCO$^+$, DCN and ND$_3$ \citep{vandishoeck2014}. In contrast, the warm conditions near protostars prompts complex chemistry that  creates large carbon chain molecules 
like  HC$_5$N, HC$_7$N, methanol (CH$_3$OH) and formaldehyde (H$_2$CO) \citep{Garrod+2008}.  Meanwhile, \add{the thermal} continuum emission from dust grains provides an indirect measure of dust and gas temperatures and densities \citep{Robitaille+2006}. Particular features in the multi-wavelength spectrum, such as the silicate feature at 9.7\,$\mu$m, indicate the underlying dust composition and size distribution \citep{Draine2003}. \add{Measuring the spectral energy density distribution (SED; $S_{\rm \nu}$) of dust emission at far-infrared (FIR; $\lambda \gtrsim 60 \mu$m) to sub-millimeter wavelengths allows one to estimate the dust mass and temperature ($T$) via the $\beta$-law given by
\begin{equation}
S_{\rm \nu} = \Omega N \kappa_0 \left( \frac{\nu}{\nu_0}^{\beta}\right) B_{\rm \nu} (T)
\end{equation}
\noindent
where $\Omega$ is the solid angle of the observing beam, $N$ is the column density, $B_{\rm \nu}(T)$ is the Planck Blackbody function, and $\kappa_0 (\nu/\nu_0)^\beta$ is the dust opacity \citep{Hildebrand1983}. $\beta$ is found to be $\sim2$ for the silicate and/or carbonaceous grains in the diffuse ISM but can deviate from this value depending on the underlying grain distribution and composition \citep{Draine1984, Draine2003}.} Additionally, polarized thermal emission from dust grains and polarized dust extinction from background stars trace the magnetic field in star-forming regions observed at (sub)millimeter and FIR wavelengths \citep{HullZhang19}

Significant progress has been made in the last decade thanks to recent wide-field long wavelength (from the infrared to radio) surveys and high-resolution sub(millimeter) observations that cover  many star-forming regions on both the large scales of $\sim$10s pc required to observe GMCs and filaments in which stars form down to the small $\sim10s$ AU size scales of circumstellar accretion disks. The advent of observatories like \emph{Spitzer} Space Telescope, Wide-field Infrared Survey Explorer (WISE), and \emph{Herschel} Space Observatory, which measure the infrared emission from thermal dust emission; and interferometers like the Berkeley-Illinois-Maryland Association (BIMA) millimeter array, the Combined Array for Research in Millimeter-wave Astronomy (CARMA), \add{Northern Extended Millimeter Array (NOEMA),} the Submillimeter Array (SMA), and the Atacama Large Millimeter/submillimeter Array (ALMA), which observe (sub)mm molecular line emission and dust continuum emission; have revolutionized our understanding of the physical processes that govern low- and high-mass star formation. In this section, we highlight many of the studies that utilize the capabilities of the aforementioned observatories and interferometers, and others, which studied both low- and high-mass star forming regions. We note that this summary is not meant to be a complete list of all recent science results or references. 

\subsection{Low-Mass Star Formation: Observations}

\begin{figure}
    \centering
    \includegraphics[width=0.85\textwidth]{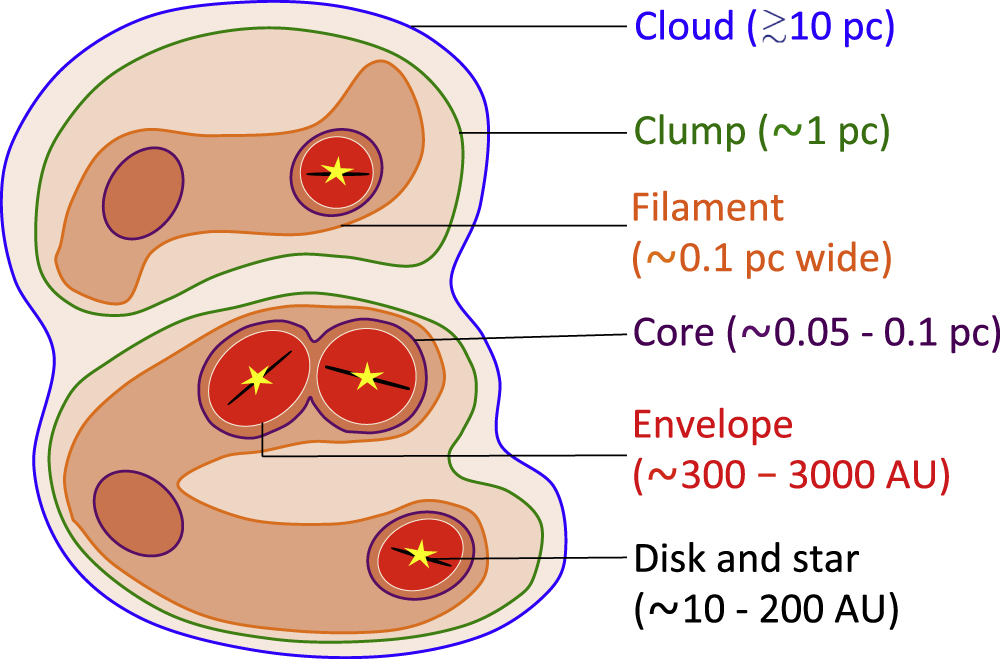}
    \caption{ A schematic of a molecular cloud showing hierarchical
structures inside the cloud responsible for low-mass star formation. The figure shows the cloud, clumps, filaments,
cores, envelopes, and protostellar systems that are discussed in this section. The image is not drawn to scale. Figure taken from \citet{Pokhrel18} \textcopyright AAS. Reproduced with permission.
    \label{fig:lmschematic}
    }
\end{figure}

Understanding the initial conditions that produce low mass stars is a key goal of current star formation studies. The birth sites of low-mass stars occur within the small, over densities of gas located within cold, dense molecular clouds (MCs). \add{MCs are turbulent with supersonic motions for size scales $\gtrsim$0.1 pc, i.e., the bulk of their volume is characterized by motions larger than their thermal sound given by $c_{\rm s} = \sqrt{kT/\mu m_{p}}$ where $T$ is the cloud temperature and $\mu$ is the mean molecular weight typically taken to be 2.33 for molecular gas at solar composition. Since they are supersonic the bulk of their density structure follows a log-normal distribution \citep[][and references therein]{Mckee2007}.} Within these clouds are networks of filaments and clumps at intermediate densities with sizes $\sim 0.1-1$ pc \citep{Andre14}. On smaller scales of $\lesssim 0.1$ pc are dense cores, the objects from which new stars are born \citep{DiFrancesco07}. When these cores become gravitationally unstable, they collapse to form one or a few young stellar objects (YSOs).  Surrounding these YSOs is the infalling core (hereafter envelope) and circumstellar disks for planets \citep[see Figure~\ref{fig:lmschematic} and ][for a summary of these scales]{Pokhrel18}. Thus, observations of star formation have the challenge to span these disparate spatial scales to connect the physical processes associated with clouds down to the young stars themselves.

Significant progress has been made in the last decade thanks to recent wide-field surveys that cover many star-forming regions on both large and small scales. In particular, many of these surveys target nearby clouds (e.g., within 500 pc), which are solely forming low-mass stars.  Many of these clouds have been grouped together into a band of recent star formation called the ``Gould Belt'' \citep[e.g.,][]{Dame01,WardThompson07b} and they offer the rare opportunity to study \emph{resolved} star formation.   

In this section, we highlight some of the major observational advances in uncovering the physical conditions of low-mass star formation, primarily due to technological advances and the aforementioned wide-field surveys.  The section is organized by spatial scales. We start with observations across entire clouds. We then discuss observations of dense cores specifically. Finally, we close on the scales of envelopes and disks.  

\subsubsection{Cloud to Core Scales}
The last decade has seen substantial improvements in observations of molecular cloud structures from filaments to individual star-forming cores. Much of these advancements have been due to legacy surveys that target most of the nearby molecular clouds at wavelengths between the infrared and radio. Here, we summarize some of the key cornerstone science results from these surveys.  

Several infrared telescopes have been used to provide entire censuses of YSOs within clouds.  Collectively, telescopes such as \emph{Spitzer} \emph{Wise}, and \emph{Herschel} covered wavelengths between $\sim 3$ \um\ to $\sim 70$ \um\ to trace YSOs by observing the warm dust surrounding them in disks and envelopes. These instruments are especially sensitive to very deeply embedded, very young YSOs that had not been  detected by the previous generation of infrared telescopes \citep[e.g.,][]{Young04,Bourke06} and a number of surveys used them to study the \emph{resolved} YSOs populations in nearby clouds \citep[e.g.,][]{Evans09,Rebull10,Megeath12,Fischer13,KoenigLeisawitz14} and to classify the YSOs into different evolutionary stages \citep[e.g.,][and references therein]{Evans09,Dunham2015}. Identifications of YSOs and their classifications are non-trivial tasks \citep[e.g.,][]{Harvey06,Hatchell07,Gutermuth09,HsiehLai13} and often require complementary data at both longer and shorter wavelengths. Subsequently, many studies combined the infrared data with complementary data to classify the YSO populations \citep[e.g.,][]{Jorgensen07,Enoch09,Stutz13,Sadavoy14} and identify several candidates for first hydrostatic cores \citep[FHSCs;][]{Chen10,Pezutto12}, a short-lived theoretical stage \citep{Larson1969} right at the onset of star formation (see Section~\ref{sec:fhsc} for more details).    

The star formation activity in a cloud appears to correlate with the quantity of dense material.  A correlation between star formation and gas surface density has been well-studied on galaxy scales \citep[see][for a review]{KennicuttEvans12}, and lately extended to individual local clouds using star counts and the masses or surface densities of the host cloud \citep[e.g.,][]{Lada10,Heiderman10,Gutermuth11,Lada13}. For local clouds, we can also construct column density probability density functions (N-PDFs) to characterize the distribution of densities within clouds. These N-PDFs show prominent high column density power-law tails for those clouds with active star formation and lognormal shapes for less active clouds \citep[e.g.,][]{Kainulainen09,Kainulainen11}. The interpretation of the N-PDF is still highly debated, with the lognormal shape often attributed to turbulence and the power-law tail attributed to gravity or pressure confinement \citep{Kainulainen11,Schneider12,Burkhart2018}. The shape of the N-PDF, however, appears to depend on the map area used in its construction \citep{Sadavoy14,Lombardi15,Alves17}, which makes theoretical interpretations of its structure more complex. Nevertheless, the N-PDF tails appear to be more robust. The N-PDF tails are primarily produced by the dense core populations in clouds \citep{HChen18} and their slopes correlate with the fraction of the youngest YSOs detected in the clouds \citep{Sadavoy14,StutzKainulainen15,Pokhrel16}.

\begin{figure}
    \centering
    \includegraphics[width=1\textwidth]{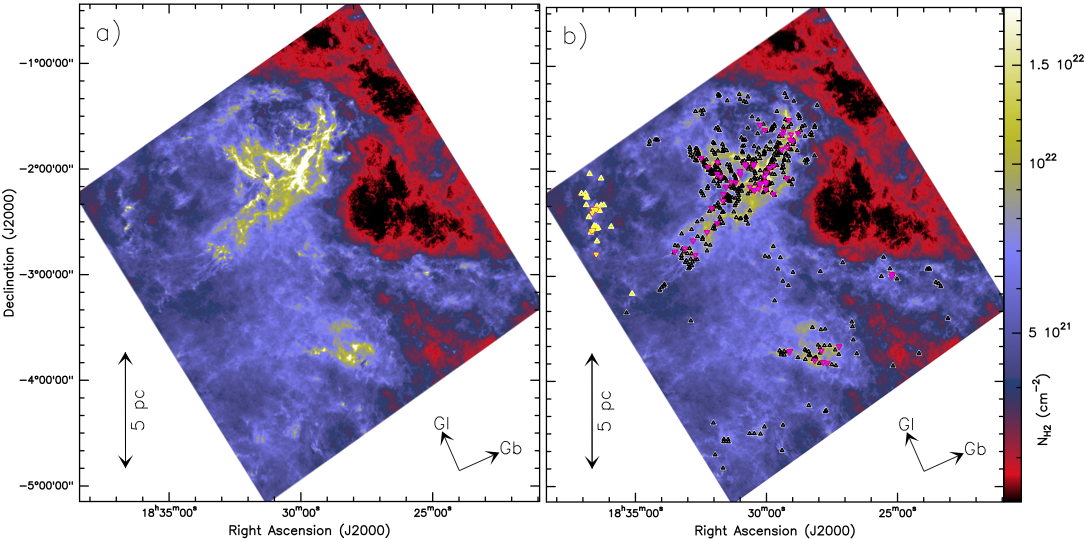}
    \caption{The prestellar and protostellar core population in the Aquila cloud complex located at a distance of $\sim$260 pc. The left panel (a) shows the H$_2$ column density map of the Aquila region at 18.2" angular resolution, as derived from \emph{Herschel} Gould Belt survey.  The right panel  (b) shows the same map as in the left panel with the positions of the 446 candidate prestellar cores and 58 protostellar cores identified in the Herschel shown as black and magenta triangles, respectively. Yellow triangles locate additional prestellar and protostellar cores which were excluded from the analysis of \cite{Konyves15} due to confusion. Figure taken from \citep{Konyves15}. Reproduced with permission \copyright ESO.}
    \label{fig:aquilacores}
\end{figure}

The dense material within clouds where stars form are generally associated with clumps and filaments. Indeed, filaments and filamentary clouds have been identified as significant to the star formation process for a number of years \citep{SchneiderElmegreen79} and observations from \emph{Herschel} and the \emph{Planck} satellite have cemented their ubiquity in the Galaxy and across star-forming clouds.  \emph{Herschel} in particular has highlighted networks of filaments within nearby and more distant clouds \citep[e.g.,][and references therein]{Andre14}. A number of studies have suggested that these elongated structures have typical widths of $\sim 0.1$ pc corresponding to their Jeans length \citep[e.g.,][]{Arzoumanian11,Palmeirim13,Arzoumanian19}, although this conclusion is still debated based on fitting techniques or resolution \citep{FernandezLopez14,Panopoulou17,Hacar18}. Nevertheless, YSOs and dense cores are found to be associated with filaments.  Observations show higher fractions of prestellar (bound) cores and YSOs toward denser filaments \citep[e.g., higher line masses; ][]{Andre10,Polychroni13,Bresnahan18,Andre19} and clusters of star formation in clumps toward the intersections of filaments \citep{Myers09,Palmeirim13,Li13,Seo19}. These observations suggest that cores and the YSOs that they host form via fragmentation processes in filaments \citep{Menshchikov10,Andre14} and that filaments can also funnel in gas to form stars and clusters \citep{Kirk13,Palmeirim13,Seo15}.  There is, however, still much discussion on the theoretical framework behind these processes in filaments. 

Another important mechanism for star formation in clouds are magnetic fields. Magnetic fields are most often inferred from dust polarization observations, but their significance is still highly debated. Nevertheless, observations show a clear connection between magnetic fields and cloud structure. Observations of polarized dust extinction from background stars \citep{Palmeirim13,HbLi13,FrancoAlves15,Cox16} or polarized dust emission from molecular clouds \citep{PlanckCollab16,Fissel16,Soler16} primarily show inferred magnetic fields that are perpendicular to dense filaments and parallel to more diffuse filaments \citep[see also,][]{PattleFissel19,HullZhang19}. Within filaments, however, the field morphologies can be more complex.  Surveys of dust polarization at far-infrared wavelengths from the Statospheric Observatory for Far-Infrared Astronomy (SOFIA) or at submillimeter wavelengths from the James Clerk Maxwell Telescope (JCMT) have revealed a wide range of field morphologies from uniform to helical \citep[e.g.,][]{Pattle17,Kwon18,Soam18,Santos19,Chuss19} on scales $<0.1$ pc that trace the interiors of filaments or down to core scales.  In addition, observations at wavelengths appear to show different field morphologies for the same cloud, indicating that dust grains are not uniformly aligned with the field \citep[see also,][]{Pattle19}. Thus, connecting observations and theoretical models of magnetic fields from the scale of the Galaxy through molecular clouds and down to the stars themselves remains a non-trivial task.

\subsubsection{Core Scales}
Cores are cold ($\sim 10$ K), compact objects with sizes of $< 0.1$ pc and densities of $>10^5$ cm$^{-3}$  that are expected to form either a single star or small stellar system when they become gravitationally unstable and collapse \citep{DiFrancesco07}. Observations of cores are often divided up into different terms, with starless cores for the ones that have not yet collapsed to form a YSO and protostellar cores for those that have a central luminous source. Starless cores can be further divided between those that are gravitationally bound and unbound. The bound (prestellar) cores are expected to be long-lived and able to collapse to form stars, whereas unbound cores are not prone to collapse. Nevertheless, there is growing evidence that some  starless cores may be pressure confined and therefore long lived \citep{Pattle15,Kirk17,HChen19}.

As the precursors for stars, dense cores have been the target of many surveys.  They are most often detected in optically thin dust emission at (sub)millimeter wavelengths \citep[e.g.,][]{Motte98,Enoch06,Jorgensen07,Konyves15,Sokol19} or with cold gas tracers \citep[e.g.,][]{Kirk07,Rosolowsky08,Friesen17,Kauffmann17} with single-dish telescopes.  Figure~\ref{fig:aquilacores} illustrates the prestellar and protostellar core population in the Aquila cloud complex \citep{Konyves15} as observed with \emph{Herschel}. A key property of dense cores is their masses. \citet{Motte98} first showed that the core mass function (CMF) resembles the stellar initial mass function (IMF) in shape, but scaled to higher masses. The  characteristic mass for CMFs is generally $\sim 1$ \Msun, which is roughly a factor of 3 higher than the stellar IMF.  As a consequence, cores are assumed to have an efficiency of 30\%\ \citep[e.g.,][]{Alves07}. This efficiency is attributed to stellar feedback such as protostellar outflows, which are collimated high-velocity gas flows emanating from YSOs that can eject entrained molecular gas from cores.\footnote{We discuss the launching of protostellar outflows in more detail in Section~\ref{sec:outflows}.} Later observations that more completely sample the \emph{prestellar} core populations of clouds found similar CMF shapes \cite[e.g.,][]{Enoch06,Sadavoy10,Pattle15,Konyves15,Marsh16}, although with efficiency factors that vary from $\sim 15$\%\ to $\sim$ 50\%\ \citep[e.g.,][]{Jorgensen08,Benedettini18}. These efficiency factors, however, assume that each core produces one star, whereas observations show that multiplicity fractions are high in YSOs (see Section \ref{env}).

Dense cores are also primarily quiescent. Many studies measuring the gas kinematics in cores have shown that low-mass prestellar cores have subsonic turbulence, whereas unbound and protostellar cores have supersonic turbulence \citep[e.g.,][]{Rosolowsky08,Sadavoy12,Friesen17,CChen19}.  Indeed, a sharp transition to coherence has been seen toward some prestellar cores using dense gas tracers \citep{Goodman1998, Pineda10,HChen19I}. In addition to the kinematics of cores, another key property is the gas chemistry. They are best probed with tracers of dense gas, although a recent survey with the IRAM 30m telescope showed that the transition critical density alone is insufficient to determine the best tracers \citep{Kauffmann17}. The reason is that cores are cold such that volatile gases like water and CO freeze out onto dust grains \citep{BerginTafalla07}. Freeze out is an important step to forming organic molecules and subsequently changes the gas chemistry in dense regions of molecular clouds and cores \citep{DiFrancesco07}. The ices can be later released back into the gas phase via outflow shocks or passive heating from YSOs.  Several surveys with \emph{Herschel} used the spectrometers to measure water toward nearby YSOs  \citep[e.g.,][]{vandishoeck11,Green13}.

For cores that host YSOs, a key observational signature are outflows, which represent gas that is entrained by a fast-moving jet and trace the spin axis of the system, the mass loss rate of the core, and the accretion rate onto the star \citep{Bally07, Dunham14b}. A number of surveys have targeted outflows across cores using $^{12}$CO observations from single-dish observations \citep[e.g.,][]{Arce10, Drabek12, Buckle12} or interferometric observations \citep[e.g.,][]{Plunkett13, Stephens17a}. A number of studies have attempted to connect outflow properties to the YSO evolutionary stage. In particular, early studies found a correlation between the outflow opening angle and the evolutionary stage \citep[e.g.,][]{Lee02, ArceSargent2006}, and also detected in more recent studies \citep[e.g.,][]{Velusamy14,Hsieh17}. The change in opening angle is attributed to mass loss in the core and less energy in the outflow itself over time. The correlation, however, is difficult to quantify as later-stage outflows are harder to identify and measure\add{,} and the outflow detection also depends on the environment density in which the gas is flowing \citep[e.g.,][]{Curtis10,Stephens17a}. Finally, several recent studies have found that outflows appear to be randomly orientated relative to the core magnetic field \citep{Hull13,Hull14} or the cloud filament elongation \citep{Stephens17a}. Alignment between YSO spin axes, magnetic fields, and filaments will have profound implications for how the star accretes material and angular momentum, affecting its evolution. It remains unclear whether or not outflows form with random orientations or end up with random orientations formed due to dynamical evolution. 

\subsubsection{Envelope and Disk Scales}\label{env}

Star formation occurs when a prestellar core gravitationally collapses to form one or more protostars. As the core undergoes inside-out collapse conservation of angular momentum causes the infalling material to form a circumstellar disk around the accreting protostar. The star-disk system is embedded within an infalling $\sim 1000$ AU envelope of dust and gas \citep[e.g.,][]{Evans2015, Tobin2018, Pokhrel18}. \add{The youngest observationally recognized protostars are classified as either Class 0/I sources based on their age and circumstellar and envelope environment. \add{For Class 0 sources, the protostar is heavily embedded by the envelope, which has a mass that is typically larger than the protostar's mass. The dividing line between Class 0/I is when the star begins to heat the surrounding dust such that there is non-trivial dust emission. This transition typically occurs at a bolometric temperature of $T_{\rm bol} \sim 70$ K or when $L_{\rm bol}/L_{\rm submm} < 0.005$ \add{\citep{Andre1994,Andre2000,Tobin20}}. The infrared excess usually indicates the presence of a circumstellar disk but their still remains a surrounding envelope. A statistical \emph{Spitzer} survey of YSOs in the Gould Belt found that the typical duration for the Class 0 and Class I phase are 0.15-0.24 and 0.31-0.48 Myr, respectively \citep{Dunham2015}}.} 

\add{Upon emergence from the envelope \add{(Class II)}, a pre-main sequence star (i.e., a low-mass star that is slowly contracting to the hydrogen-burning main sequence) surrounded by circumstellar dusty disk remains after the surrounding envelope dissipates and this disk is the site for planet formation \citep{Tobin20}. \add{These sources are typically known as classical T Tauri stars.} Finally, Class III sources are pre-main sequence stars that are no longer accreting significant amounts of matter and are known as weak-lined T Tauri stars \citep[][and references therein]{Mckee2007}.}

High-resolution, interferometric observations with the BIMA, CARMA, \add{NOEMA}, SMA, and ALMA interferometers have revolutionized the study of the environments of low-mass protostars from the infalling core envelope scales of several $\times$ 1000 AU down to disk scales of a few $\times$ 10 AU \citep[see ][and references therein]{HullZhang19}. Given the close proximity of numerous low-mass star forming regions, several studies have statistically constrained how stars gain their mass by analyzing fragmentation, disk and envelope evolution, angular momentum transport, and the outflow energetics of numerous class 0/I sources. For example, polarization studies have found that magnetic fields play a role in regulating the infall of material all the way down to the $\sim$1000 AU scales of envelopes and that feedback from outflows may alter the magnetic field morphology \add{\citep[e.g.,][]{Hull14, Hull+2017, Hull2017b, Maury2018}}. \add{On smaller scales, high angular resolution studies, including polarization studies, have observed small $\lesssim10s$ AU disks and/or streamers that surround low-mass protostars and proto-binaries  \citep{Sadavoy2018a, Sadavoy2018b, Alves2019, Tobin20}.} Additionally, measured disk masses and envelopes of Class 0/I sources \add{find} that the disk mass remains roughly constant between Class 0 and Class I sources while the envelope mass tends to decrease over time \citep[e.g.,][]{Stephens18, Andersen2019}. Furthermore, disk formation likely occurs rapidly during the \emph{early} Class 0 phase for low-mass YSOs \citep{Gerin2017, Andersen2019}. The envelopes are depleted by both accretion onto the star-disk system and by ejection due to energetic outflows that drive out entrained molecular material \citep[e.g.,][]{ArceSargent2006, Koyamatsu2014, Yang2018}. \add{Observations demonstrate that} the kinematic signatures of \add{protostellar} envelopes have higher line widths than the typical line widths found in Perseus at the core and filament scales suggesting that gas infall and feedback from outflows increases the envelope energetics \citep{Stephens18}. 

\add{Additionally, infrared and submillimeter studies have also detected episodic accretion onto \add{young Class 0/I} protostars. Episodic accretion is described by a series of relatively brief but dramatic spikes in the accretion rate over the star formation period. These variations causes luminosity changes, typically above $\sim10-20\%$ of the baseline luminosity, that are reprocessed by the surrounding envelope. Studies have found that the accretion variability can last as short as a $\sim$few weeks to several years \citep[e.g.,][]{Billot2012, Safron2015, Mairs2017}. Such bursts may be triggered by disk fragmentation due to gravitational instability, leading to brief but higher accretion rates.}

Numerous studies have found that the multiplicity of low-mass YSOs is common with Class 0 sources exhibiting a higher multiplicity fraction than Class I sources \citep{Chen2013, Tobin+2016, Sadavoy2017a, Tobin2018b}.  \add{\citet{Chen2013} found that 64\% of Class 0 protostars, that are located in nearby  ($d \lesssim$ 500 pc) molecular clouds are in multiple systems with separations ranging from 50 AU to 5000 AU, whereas this fraction decreases by a factor of $\sim$2 for Class I sources and by a factor of $\sim$3 among main-sequence stars, with a similar range of separations.} Companion YSOs can form via core or disk fragmentation. With core fragmentation wide companions ($d \gtrsim 1000$ AU) can form via Jeans fragmentation, in which fragmentation is driven entirely by the competition between gravity and thermal support, turbulent core fragmentation or by rotationally induced fragmentation \citep{Offner+2009, Chen2010, LeeDunham2015, Pokhrel18}. Closer in companions ($d \lesssim 100$ AU) can be formed via disk fragmentation induced by gravitational instabilities in a disk \citep[e.g.,][]{Tobin2013, Tobin2018b, Alves2019}. 

The presence of companions also affect disk sizes and lifetimes: YSO and young star systems with a close-in stellar companion have a lower fraction of infrared-identified disks than those without such companions, indicating shorter disk sizes and lifetimes in close multiple systems \citep{Chen2013,Kounkel2019,Tobin20}. \add{Additionally, studies also find that disk size and mass decrease with protostellar age. \citet{Tobin20} used high-resolution ALMA and VLA observations to measure the dust disk radii and masses towards a large sample of protostars in Orion and found that the disk size and mass (as measured by the dust emission) decreases with evolutionary age: the mean dust disk radii are $\sim45$ and $\sim37$ au for Class 0 and Class I protostars, respectively; and that the protostellar disk mass is typically a factor of $\gtrsim4$ larger than the dust masses for observed in Class II disks that surround more evolved pre-main sequence stars. Their results suggest that planet formation may need to at least begin during the protostellar phase.}

\subsection{High-mass Star Formation: Observations}
\label{sec:hmsfo}

Observations of high-mass star formation are hindered by the rarity and consequent distance of forming high-mass stars. They form in clustered environments that are characterized by higher surface densities and larger velocity dispersions than nearby low-mass star forming regions \citep[e.g.,][]{TanPPVI+2014, Zhang2015a, Liu2018a}. 
However, it is still highly debated if high-mass star formation is simply a scaled up version of low-mass star formation in which massive stars form via the monolithic collapse of massive prestellar cores that are supported by turbulence and/or magnetic fields rather than thermal motions \citep{McKee+2003, TanPPVI+2014} or \add{if they form via larger scale accretion flows due to gravity or converging, inertial flows that naturally occur in supersonic turbulence from the surrounding molecular cloud \citep{Bonnell+2001, VazquezS+2003, Padoan2019}.} 

The former scenario, known as the Turbulent Core (TC) model, requires that massive prestellar cores are in approximate virial equilibrium supported by turbulence and/or magnetic fields and these cores become marginally unstable to collapse to form a massive star or massive multiple system. The resulting formation timescale is several times the core freefall timescale ($t_{\rm ff} \lesssim 10^5 \, \rm{yr}$) and the high degree of turbulence causes clumping, resulting in high accretion rates ($\dot{M}_{\rm acc} \sim 10^{-4} \msun \rm{yr^{-1}}$) that can overcome feedback associated with the star's large luminosity \citep{McKee+2003}. In this scenario, the core represents the entire mass reservoir available for the formation of a single massive star or a massive multiple system, since on larger scales, the cloud is simultaneously supported and fragmented by turbulence \citep{VazquezS+2003}.

The latter scenario, known as Competitive Accretion (\add{CA}), instead posits that low-mass protostellar seeds will accrete unbound gas
within the clump as determined by its tidal limits, and when they
become massive enough, they will then accrete at the Bondi-Hoyle accretion rate, $M_{\rm BH} \propto v^{-3}$ where $v$ is the relative velocity of the gas. In the CA model, accretion is favored toward the center of the gravitational potential and therefore stars near the center of the cluster gain the most mass. This model achieves high accretion rates onto the protostar under subvirial initial conditions, in contrast to the virialized conditions of the turbulent core model, since the gas velocity dispersion is low. This model instead forms a cluster of stars with varying masses and therefore high-mass star formation is closely linked to cluster formation. Additionally, it has also been suggested that molecular clouds may be in a regime of global hierarchical collapse (GHC), in which all size scales are contracting gravitationally, and accreting from the next larger scale \citep{VazquezS+2019}. In this scenario, the nonthermal motions in molecular clouds and their substructures (filaments, clumps, and cores) may consist of a combination of infall motions and truly turbulent motions \citep{BP+2011, BP+2018,VazquezS+2019}. The GHC scenario allows for large scale accretion flows to directly fed high-mass star forming regions. \add{\citet{Padoan2019} argue that instead of gravitational collapse the large scale accretion flows that are required to directly fed high-mass star forming regions are instead supplied by the large-scale inertial flows driven by supersonic turbulence within the cloud. They refer to this as the \textit{Inertial-inflow model} of high-mass star formation.}

Dust emission and molecular tracers are useful tools to study the physical conditions such as the temperature, density, and velocity structure within massive clumps that are the sites of high-mass star formation. Given the recent advances in long-wavelength (from the infrared to radio) and interferometric surveys, we can now test the aforementioned theories directly to determine how high-mass stars form. Such observations have mapped the evolutionary sequence of high-mass star formation across the galaxy from quiescent non-star forming clumps to active star forming clumps that host ultra compact \hii\ (UC\hii) regions and the larger filamentary complexes in which high-mass stars form. Additionally, ALMA has enabled us to extend observations of high mass YSOs (HMYSOs) to much greater distances, and therefore larger samples, than ever before.  In particular, ALMA's long baselines allow us to probe the inner $\lesssim 1000$ AU toward HMYSOs that are actively accreting. In this section, we highlight some of the recent major observational advances in studying high-mass star formation from the size scales of clouds and clumps ($\sim 1-10$\add{s pc}) down to the size scales of protostellar envelopes and disks ($\sim 10s-1000s$ AU). 

\subsubsection{Cloud to Clump Scales}
Massive stars form in dense ($10^4-10^6 \; \rm{cm^{-3}}$), cold turbulent gas within GMCs and giant massive filaments \citep[e.g.,][]{Zhang2015a,Li+2016, Lin+2019, Urquhart2018, Zhang2019}. Within these dense clouds are condensations commonly referred to as clumps with masses of a $\sim10s - 10^4 \; \msun$ \citep[e.g.,][]{Schuller2009, Urquhart2018}. These clumps are generally subdivided in two groups: quiescent (starless) and star forming clumps that are undergoing active star formation. Large scale surveys, like the the APEX telescope large area survey of the galaxy (ATLASGAL) 850 $\mu\rm{m}$ survey, have mapped the distribution of massive star forming regions in the galactic disk. \cite{Urquhart2018} performed a statistical analysis on $\sim 8000$ dense clumps observed by ATLASGAL. They found that dense clumps that are capable of or actively forming high-mass stars have a mean size of $0.72 \pm 0.01$ pc and primarily trace the dense gas in the spiral arms of the Galaxy. This survey also found that the vast majority of clumps ($\sim88\%$) are undergoing active star formation at different evolutionary stages, suggesting that star formation in dense clumps occurs rapidly. Their result suggests that the clumps build themselves rapidly and therefore global infall from the surrounding cloud likely does not drive star formation on the large clump scale, as predicted by the GHC model. Clumps undergoing active star formation are significantly more centrally condensed and spherical in shape as compared to quiescent clumps indicative of gravitational collapse \citep{Urquhart2015, Urquhart2018}. These active clumps also exhibit higher velocity dispersions and temperature gradients, which is likely a result of heating by stellar feedback. 

\begin{figure}
    \centering
    \includegraphics[width=1\textwidth]{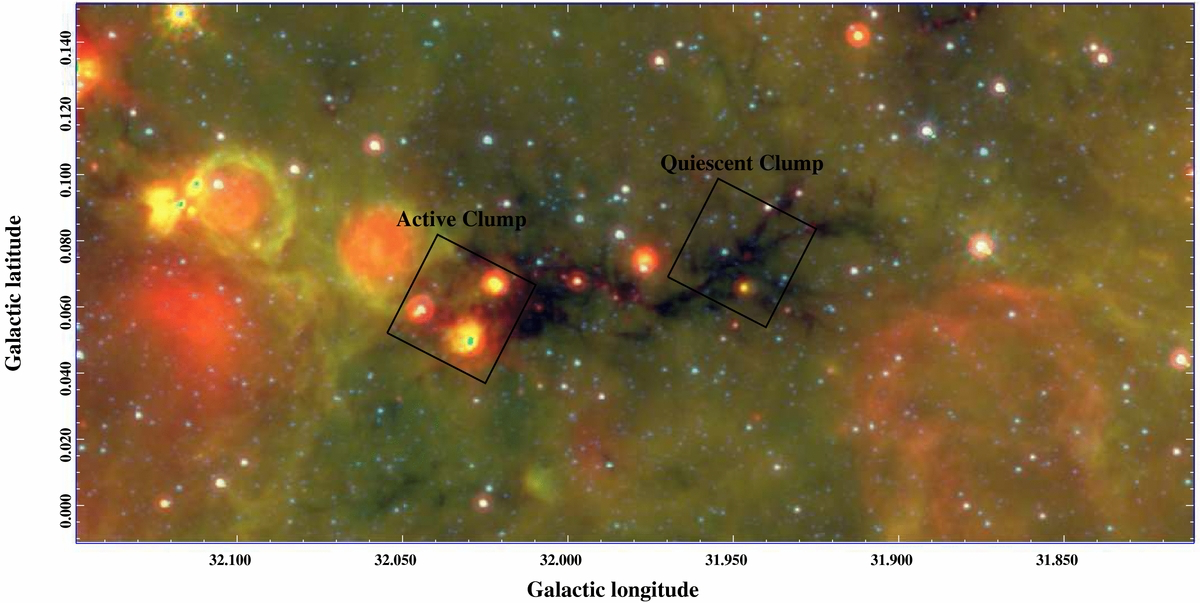}
    \caption{The G32.02+0.06 IRDC that is embedded within a massive molecular filament. This IRDC shows different stages of massive star formation, from extended \hii\ regions showing active star formation to infrared dark, cold gas. The two black squares highlight the active (left) and quiescent (black) clumps. The background is a \emph{Spitzer} three-color image (red: MIPSGAL 24$\mu$m, green and blue: GLIMPSE 8$\mu$m and 4.5$\mu$m). Figure taken from \cite{Battersby2014} \textcopyright AAS. Reproduced with permission.}
    \label{fig:battersby}
\end{figure}

Massive star forming clumps are typically embedded in larger structures known as infrared dark clouds (IRDCs), which are the dense precursors to stellar clusters and are dense molecular clouds seen as extinction features against the bright mid-infrared  Galactic background \citep[e.g.,][]{Rathborne2006, Rathborne2010, Battersby2014, Contreras2018}. A key signature of active star formation in IRDCs and clumps is an excess of 24-70 $\mu$m emission, indicating that they contain one or more protostars \citep[e.g.,][]{Rathborne2010, Pillai+2019}. To illustrate these features, we show the G32.02+0.06 IRDC that is embedded within a massive galactic filament in Figure~\ref{fig:battersby}. This IRDC contains both an active star forming region that is infrared bright showing active \hii\ regions and a quiescent clump that is infrared dark denoting a lack of star formation.

Observations of quiescent clumps suggest cloud collapse and high-mass star formation occurs above a density threshold. \cite{Traficante+2018, Traficante2020} studied the dynamics of quiescient (70 $\mu$m dark) clumps with varying surface densities and found that the dynamics of non-star forming clumps with high surface densities in excess of $\Sigma \gtrsim 0.1 \; \rm{g \, cm^{-2}}$ are mostly gravity-driven rather than turbulence-driven and are in a state of global gravitational collapse, and are therefore likely the precursors to high-mass stars.  The rate of collapse of these clumps can be measured by the optically thin N$_2$H+(1-0) line for the ``blue asymmetry" spectroscopic signature of infall motion given by $\displaystyle\frac{I_{\rm blue}-I_{\rm red}}{I_{\rm blue}+I_{\rm red}}$, where $I_{\rm blue}$ ($I_{\rm red}$) is the blue-shifted (red-shifted) integrated line intensity. Using this diagnostic for a large sample of massive clumps observed with the Millimetre Astronomy Legacy Team 90 GHz (MALT90) Survey, \cite{Jackson2019} found that the clumps are predominantly undergoing gravitational collapse and that the rate of collapse is larger for the earliest evolutionary stages (quiescent, protostellar, and UC\hii\ region) than for the later \hii\ and photodissociation region classifications. Hence, these results suggest that as star formation and therefore stellar feedback becomes significant, the energy injection by feedback likely reduces the rate of gravitational collapse. 

Strong protostellar outflows from HMYSOs are also a signature of active star formation. Outflows are typically traced by the high-velocity entrained gas observed with molecular line (usually CO) emission. For massive protostars, outflows are thought to be a scaled-up version of the accretion related outflow-generation mechanism associated with disks and jets in low-mass YSOs and the outflow mass-loss rate is tightly correlated with the accretion rate onto the protostar. \add{By assuming the outflow mass-loss rate is $\sim10-30\%$ of the accretion rate, s}tatistical studies of the outflow mass-loss rates in high-mass star forming regions outflows infer mass accretion rates of $\sim 10^{-4}-10^{-3} \; \msun \, \rm{yr^{-1}}$, indicative of high-mass star formation and in agreement with the TC and CA models \citep[e.g., ][]{Maud2015, Yang2018, LiZhou2019}. \cite{Yang2018, LiZhou2019} found that the rate of detection of outflows increases with evolutionary stage (e.g., form the protostellar to the HII region stage) and the outflow energetics in these clumps are dominated by the most massive and luminous protostars.

\subsubsection{From Clump to Core Scales}
On smaller scales, IRDCs and massive clumps fragment into prestellar cores ($r\sim 0.1$ pc) with masses between $\sim 10-100 \; \msun$. These cores likely result from turbulent fragmentation since they have masses larger than the mass and length scale dictated by thermal Jeans fragmentation \citep{Zhang2009, LuZhang2015}. They tend to be embedded in filamentary structures within the clouds that can span several parsecs in length, or at the sites where several filaments converge, termed hubs \citep{Battersby2014, Peretto2013, Henshaw2017, Trevino2019, Pillai+2019}.  The filaments themselves accrete from the cloud scale, potentially feeding the cores and subsequent protostars \citep{Peretto+2014, Tige2017, Contreras2018, Williams2018, LuZhang2018, Trevino2019, Russeil2019}. In this case, the mass reservoir for star formation in the hubs extends at least to the clump (pc) scale, favoring the GHC and CA scenarios.

Most studies have found that massive cores \add{are} supersonic but are typically subvirial (i.e., not supported by turbulence) and should collapse within a gravitational freefall time if the cores are not supported by magnetic fields \citep[e.g.][]{Kauffmann2013, Battersby2014, Contreras2018, Kong2018}. These studies suggest that strong magnetic fields of the order of $\sim$1 mG are required for stabilizing massive prestellar cores. In a few cases, fields this strong have been measured in high-mass star forming regions and massive cores \cite[see][ and references therein]{HullZhang19}. \add{More measurements of the magnetic field strength of massive prestellar cores can help determine the demographics and stability of massive prestellar cores, potentially supporting the TC model.} 

Observations of massive cores show that some further undergo thermal Jeans fragmentation \citep[e.g.,][]{Palau2015, Beuther2019, Sanhueza2019} while others do not \citep[e.g.,][]{Battersby2014, Csengeri2017, Louvet2019}. These fragments may be the precursers of low-mass prestellar cores that can accrete from the surrounding unbound gas to form massive stars as described by the CA model. However, magnetic fields may provide additional pressure support and help regulate the rate of collapse and amount of fragmentation \citep{Fontani+2016,Csengeri2017}. Likewise, radiative feedback from the first-formed high-mass protostar within the core reduces fragmentation and promotes the formation of a system with a few, higher mass stars rather than a cluster of low-mass stars (see Section~\ref{sec:SFtheory}). 

Whether high-mass and low-mass star formation occurs coevally is still debated but has implications for high-mass and star cluster formation. \cite{Pillai+2019} observed two well-studied IRDCs, G11.11-0.12 and G28.34+0.06, with the SMA. These IRDCs appear starless because they are dark at 70-100 $\mu$m. They found that the dense clumps within these IRDCs have fragmented into several low- to high-mass cores within the filamentary structure of the enveloping cloud. Furthermore, they detect high-velocity CO 2-1 line emission indicative of compact outflows suggesting that these clumps are undergoing active low-mass and possibly early high-mass star formation. Their results suggest that low-mass stars might form first or coevally with high-mass stars during the youngest phase ($<$0.05 Myr) of high-mass star formation.

The numerous studies discussed above suggest a dynamical scenario of high-mass star formation in which massive cores are built by accreting gas from the surrounding clump rather than fragmentation processes alone. In agreement with this scenario, \cite{Contreras2018} studied a highly subvirial, collapsing massive prestellar core with mass $17.6 \; \msun$ that is heavily accreting from its natal cloud at a rate of $1.96 \times 10^{-3} \; \msun \; \rm{yr^{-1}}$ that has not fragmented and shows no evidence for outflows. The low-level of fragmentation and result that the core is $\sim6$ times the clump's Jeans' mass suggests that this core is in an intermediate regime between the TC and CA models. This finding suggests that massive core and star formation may precede simultaneously as in the GHC scenario. However, a more statistical sample measuring the dynamics and growth of massive prestellar cores is required to test this theory.  

\subsubsection{Envelope and Disk Scales}
Whether high-mass stars form from the collapse of high-mass prestellar cores (TC model) or from inflow from larger scales (CA, GHC, and Inertial-inflow models) still remains an open question. However, numerous high-resolution ($\sim 1000$s AU scale) observations of accreting HMYSOs suggest that high-mass stars form similarly to their low-mass counterparts via infall from a surrounding envelope and the development of an accretion disk that can provide an anisotropic accretion flow onto the star (see Section~\ref{sec:SFtheory} for more details).

Evidence of infall from protostellar envelopes onto HMYOs has been detected for a large number of sources \cite[e.g.,][]{Fuller2005, vanderTak2019}. \cite{Fuller2005} detected infall onto 22 $850\mu$m continuum sources believed to be candidate HMYSOs. These sources showed significant excess of blue asymmetric line profiles for several molecular line species, suggesting that the material around these high mass sources is infalling at rates of $2\times10^{-4}-10^{-3} \; \msun \; \rm{yr}^{-1}$. Similarly, \cite{vanderTak2019} measured velocity shifts between the H$_2^{18}$O absorption and C$^{18}$O emission lines \add{from data taken with the HIFI instrument on \textit{Herschel}} for 19 HMYOs at different evolutionary phases to measure the infall motions in their surrounding envelopes. They concluded that infall motions are common in the highly embedded phase of HMYOs, with typical accretion rates of $\sim10^{-4} \; \msun \; \rm{yr}^{-1}$. Furthermore, consistent with the TC model, they find that the highest accretion rates occur for the most massive sources and that the accretion rates may increase with evolutionary phase. 

The infalling material will circularize  to conserve angular momentum as it falls to the star and forms a Keplerian accretion disk if the magnetic field is not strong enough to \add{transfer} a significant amount of angular momentum \add{from small to large scales, an effect termed as magnetic braking}. \add{If magnetic fields are relatively ordered then they will remove angular momentum from the accretion flow and the material will be circularized closer to the star. Regardless, t}he location at which the infalling material circularizes is known as the centrifugal barrier. Using high angular resolution ALMA observations, \cite{Csengeri2018} reported the first detection of the centrifugal barrier at a large radius of 300-800 AU around a high-mass 11-16 $\msun$ protostar ($L_{\rm bol} = 1.3\times 10^4 \lsun$) surrounded by a massive core of $\sim 120 \; \msun$. \add{They suggest} that the indication for \add{an accretion disk with a radius $\gtrsim$500 au predicts that magnetic braking has not sufficiently transported angular momentum from smaller to larger scales for the infalling gas, suggesting the magnetic field in the collapsing envelope is weak and/or disordered for this object}. They also find that the core in which the HMYSO is embedded in does not show fragmentation and appears to be collapsing monolithically consistent with the TC model.

The result of \cite{Csengeri2018} is the only study thus far to demonstrate how the infall of envelope material can build an accretion disk around accreting HMYOs. However, high-angular ALMA observations have reported the presence of disks, either showing \add{(roughly)} Keplerian \add{or slowly rotating motion}, around several HMYSOs and some of these disk sizes agree with the large radius inferred from \cite{Csengeri2018} \add{whereas some have small radii, suggesting that the material was circularized closer to the star due to magnetic braking as predicted by numerical simulations \citep{Matsushita2017, Kolligan2018}. We refer the reader to the review by \citet{Zhao2020} that describes magnetic braking in disk formation around young stellar objects.}  A list of reported disks is given in Table \ref{tab:hmysodisk} \add{and the diversity of disk sizes suggest that magnetic fields are dynamically important in high-mass star and disk formation}. The key feature of well-measured disks around HMYOs is that they contain masses smaller than the central star, which is not surprising given that more massive disks would be highly Toomre unstable \citep{Ahmadi2019a}. However, recent observations have shown that such disks can fragment and form low-mass companion stars that may eventually grow in mass via disk accretion \citep{Ilee2018b,Zapata2019}.

Disks are likely present in earlier stages, but they are more difficult to detect because of the high optical depths of the surrounding material \citep[e.g.,]{Krumholz2007a}. They may also be smaller in both mass and radius because they are externally disrupted by the ongoing accretion flow \citep{Goddi2018a}. Several cases of confirmed forming high-mass stars have disk size upper limits $<500$ AU \citep[e.g.,][]{Maud2017a,Goddi2018a}. In these cases, there is still strong evidence for the presence of disks, since powerful outflows are observed.

\begin{figure}
    \centering
    \includegraphics[width=0.32\textwidth]{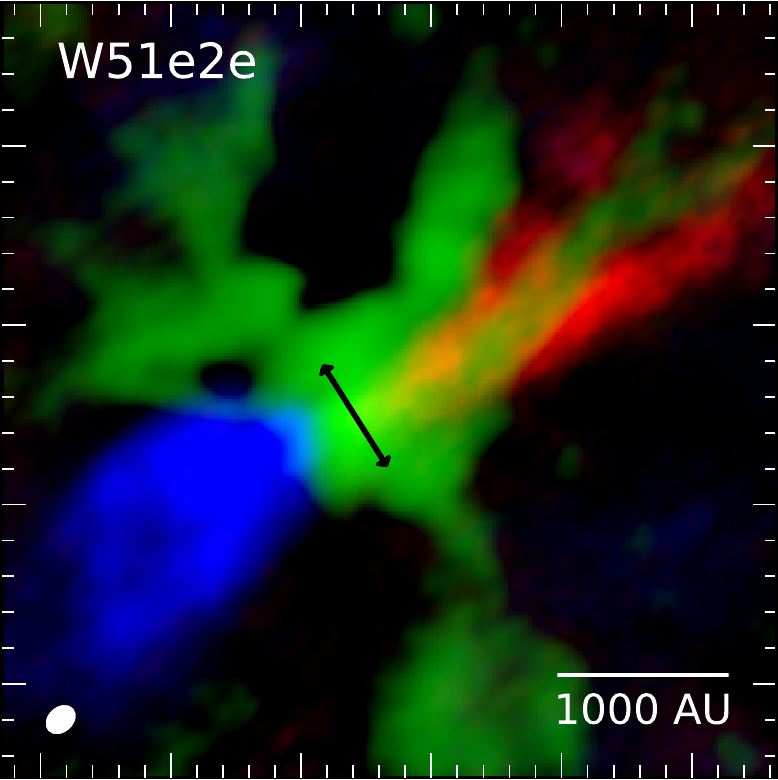}
    \includegraphics[width=0.32\textwidth]{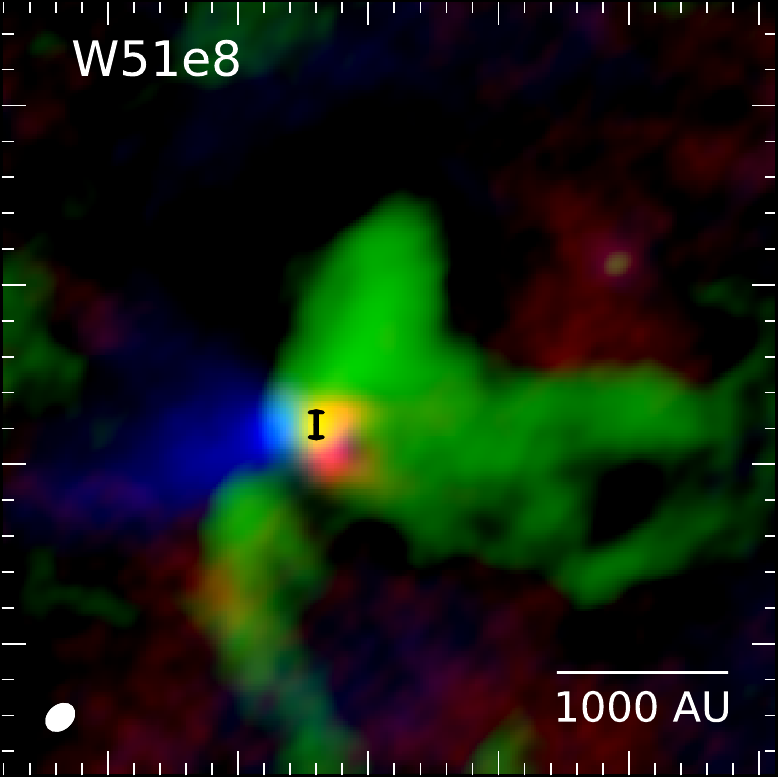}
    \includegraphics[width=0.32\textwidth]{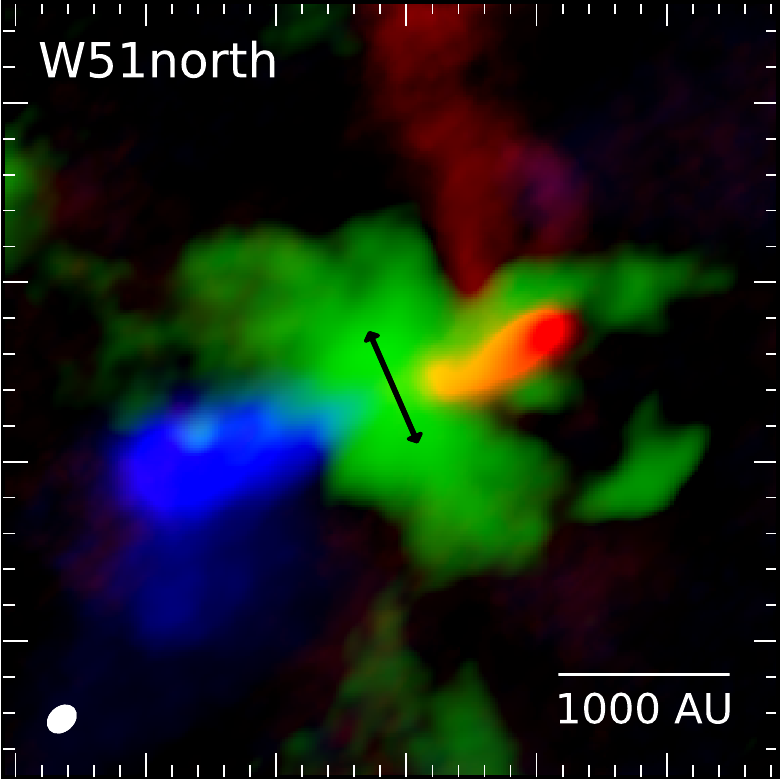}
    \caption{Figures showing the lack of observed disks (but hints at their existence) in W51; green shows continuum, red and blue show red and blueshifted SiO J=5-4 outflows.  
    This figure is reproduced from \citet{Goddi2018a}.}
    \label{fig:w51nodisk}
\end{figure}

In addition to disk accretion, HMYSOs are also likely fed material through filaments. High-angular resolution ALMA observations have determined that the environments around HMYSOs on $\sim$100-1000 AU scales is highly chaotic and filamentary \citep{Maud2017a,Goddi2018a}. Figure~\ref{fig:w51nodisk} shows three observed highly embedded HMYSOs in W51, a high-mass star forming complex located at a distance of $\sim$5.4 kpc. Here, the green shows the continuum emission and the red and blue show the red and blue shifted  SiO  J=5-4 emission that traces outflows that are likely driven by a small, unresolved disk. The green continuum emission for these sources show that accretion onto the HMYOs is predominantly asymmetric and disordered suggesting that filamentary streamers can also deliver material to the central sources.  The multi-directional accretion channels may inhibit the formation of a large, steady disc during the early highly-embedded phase of high-mass star formation.


\begin{table}[]
    \centering
    \caption{Disk or disk like structures properties around HMYSOs}
    \begin{tabular}{c|c|c|c|c}
    Object & Star Mass & Disk Mass & Disk Radius  & References \\
           & \msun     & \msun     & au           & \\
    \hline
    Orion Source I & $15\pm2$ & $<0.2$ & 75-100 & \cite{Plambeck2016a,Ginsburg2018b}\\
    G17 & $45 \pm 10$ & $<2.6$ & 120 &  \cite{Maud2018a,Maud2019a} \\
    G16 & $10\pm2$ & $1.8\pm0.3$ & 500 & \cite{Moscadelli2019a} \\
    G20 & 20  & 1.6 & 2500 &  \cite{Sanna2018a} \\
    AFGL 4176 & 20 & 2--8 & 1000 & \cite{Johnston2015a,Sanna2018a} \\ 
    S255IR NIRS3 & 20 & 0.3 & 500 &  \cite{Zinchenko2015a, Caratti-O-Garatti2016a} \\ 
    G11.92-0.61 MM1 & $34\pm5$ & 2.2-5.8 & 480 & \cite{Ilee2016a,Ilee2018b} \\ 
    IRAS 23033+5951 MMS1b & $18.8\pm1.6$ & -- & -- & \cite{Bosco2019a} \\  
    GGD27 MM1 & 18 & 4 & 300 AU & \cite{Girart2017a, Girart2018a} \\ 
    G35.20-0.74N & 18 $\pm$ 3 & 3 & 2500 & \cite{Sanchez-Monge2013a,Sanchez-Monge2014a} \\
    IRAS20126+4104 & 12 & 1.5 & 860 & \cite{Cesaroni2014a,Chen2016e} \\
    IRAS16547-4247 & 20 & 4 & 870 & \cite{Zapata2015a,Zapata2019} \\
    \end{tabular}
    \label{tab:hmysodisk}
    When mass error bars are not given, the measurements should be taken as loose estimates, e.g., based on consistency
    checks between a stellar type and the upper-limit luminosity.
    Radius estimates are wavelength-dependent.
\end{table}


\section{Analytical and Numerical Modeling in Low-Mass and High-Mass Star Formation}
\label{sec:SFtheory}

As discussed above, low- and high-mass star formation is a multi-scale, multi-physics problem ranging from the size scale of several pc down to sub-AU scales. Multi-wavelength observations and surveys from the radio to near-infrared with observatories like  \textit{ALMA, Herschel,} and \textit{Spitzer} have shed light on the star formation process, however a more intuitive and physical picture has been elucidated with recent theoretical and numerical work. 

Modern theories tend to view star formation as a continuum, where physical processes, such as radiation pressure or turbulence, are more or less important for different stellar masses and size scales. Despite growing computing power, however, it remains prohibitively expensive to model star cluster formation from super-pc to sub-au scales. Consequently, it remains informative to treat the formation of individual stars as isolated events to study the microphysics and physical complexity on sub-pc to sub-au scales.

Here we provide the theoretical background of low- and high-mass star formation and highlight the recent advances numerical simulations have provided in understanding these processes. In this section, we first summarize analytic models for isolated core-collapse, which have historically provided the foundation for understanding the relationship between thermal pressure, magnetic fields, turbulence, and gravity. Next, we summarize the numerical methods currently used to perform detailed numerical simulations of star formation. In what follows, we review the recent advancements in understanding the physical complexities involved in low- and high-mass star formation with numerical simulations.

\subsection{Analytical Core Collapse Models and Characteristic Physical Parameters}

\label{sec:corecol}
A small set of conditions undergoing  gravitational collapse are amendable to analytic solution.  The simplest case -- the collapse of an infinite uniform, isothermal medium -- was first worked out  independently by Larson and Penston \citep{Larson1969,Penston1969}. However, a slightly more realistic configuration occurs if the gas is centrally condensed. If the density initially spans at least a couple orders of magnitude, then the solution limits to the collapse of an isothermal sphere  \cite{Chandrasekhar1939}. In this limit, the gas density is given by
\begin{equation}
\rho = \frac{c_s^2}{2 \pi G r^2},  \label{eq:isoden}  
\end{equation}
where $c_s$ is the thermal sound speed. While this implies somewhat unnaturally that the density is infinite at the center, the collapse solution is conveniently self-similar. The resulting density and velocity distributions are scale free and have no characteristic density. One additional feature of this configuration is that the core undergoes an inside-out collapse, during which the gas remains isothermal. Once collapse begins the density profile approaches the free-fall form of $\rho \propto r^{-3/2}$.

Collapse including some initial slow rotation\add{,  with rotation rate $\Omega$,} follows a similar analytic solution, where the outer rotating envelope density distribution is comparable to that of equation \ref{eq:isoden}. Rotation naturally allows for the formation of a disk inside the centrifugal radius, \add{$R_c = \Omega^2 R^4/GM$  or in terms of $c_s$ this becomes $R_c = G^3M^3\Omega^2/16 c_s^8$}, where $\rho \propto r^{-1/2}$ due to conservation of angular momentum of the infalling material \citep{Terebey+1984}. Meanwhile, the collapsing gas at intermediate radii limit to the infall profile
\begin{equation}
\rho = \frac{\dot M}{4 \pi (2G M)^{1/2}r^{3/2}},
\end{equation}
where $\dot M$ is the accretion rate.

These initial conditions, while overly simplistic, have provided the basis for deriving the characteristic timescales, masses, and accretion rates of low-mass star formation since prestellar cores that form low-mass stars are subsonic and roughly isothermal. Namely,
\begin{equation}
t_{\rm ff} = \sqrt{\frac{3 \pi}{32 G \rho}} = 0.44 \left(\frac{n_{\rm H} }{10^4\, {\rm cm}^{-3}}\right)^{-1/2} {\rm Myr}, \label{eqn:tff}
\end{equation}
the free-fall time for the gravitational collapse of a pressureless gas,
\begin{equation}
\dot M = 0.975 \frac{c_s^3}{G} = 1.5 \times 10^{-6} \left(\frac{T}{10\, {\rm K}} \right)^{3/2} \msun {\rm yr}\e, 
\label{eqn:mdot}
\end{equation}
the fiducial infall/accretion rate of an isothermal centrally condensed sphere \citep{Shu1977}, and
\begin{equation}
M_{\rm BE} = 1.18 \frac{c_s^4}{\sqrt{G^3 P_s}} = 3.6 \left( \frac{T}{10\, {\rm K}} \right)^{2} \left(\frac{P_s /k}{10^4\, {\rm K\, cm}^{-3} }\right)^{-1/2} \msun, \label{eqn:BE}
\end{equation}
the maximum stable mass of a sphere of gas confined by pressure, $P_s$, or the ``Bonnor-Ebert mass" \citep{Bonnor1956}.
In the presence of magnetic fields, the characteristic critical stable mass becomes $M_{\rm cr} = M_{\rm BE} + M_\phi$ \citep{McKee1989}, where $M_\phi$ is the mass at which gravitational collapse is prohibited by magnetic pressure support,
\begin{equation}
M_{\rm \phi} =  0.13 \frac{\phi}{G^{1/2}} = 0.75 \left(\frac{R}{0.1\, {\rm pc}}\right)^2 \left( \frac{B}{10\, \mu{\rm G}} \right) \msun,    
\end{equation}
where $\phi = \pi R^2 B $ is the magnetic flux of a sphere with uniform field $B$ \citep{MouschoviasSpitzer1976}. Then the degree to which a given mass is supported by magnetic fields is  quantified by the mass-to-critical flux ratio, $\mu_\phi = M/M_{\rm \phi}$. 

Since the seminal study of cloud linewidths by Larson in 1981 showed a correlation between velocity dispersion and spatial scale \citep{Larson1981}, the impact of non-thermal velocities has been central to many star-formation models. Turbulence is an intrinsically non-linear and multi-scale process, which is not amenable to simple analytic description. Consequently, turbulence is often treated as a non-thermal, isotropic pressure and normalized according to the gas sound speed, i.e., $P_{\rm NT} = \rho \sigma_{\rm NT}^2$, where $\sigma^2 = c_s^2 + \sigma_{\rm NT}^2$, $\sigma$ is the 1D gas velocity dispersion and $\sigma_{\rm NT}$ is the non-thermal velocity component. This naturally suggests that equations \ref{eqn:mdot} and \ref{eqn:BE} can be modified by substituting the effective velocity dispersion for the sound speed. The amount of turbulence can be parameterized by the gas Mach number, a scale-free parameter normalized by the thermal velocity:
\begin{equation}
\mathcal{M} = \frac{\sigma_{\rm NT}}{c_s}.   
\end{equation}
Observationally, low-mass cores are characterized by sub-sonic velocity dispersions ($\mathcal{M} < 1$) \add{\citep[e.g.,][]{Barranco1998a, Hacar2011}}.

In the context of high-mass star formation, massive clumps and cores with typical masses of $\sim 10~\msun-100~\msun$, which may be the birthsites of high-mass stars as discussed in Section~\ref{sec:hmsfo}, are supersonic ($\mathcal{M} > 1$) with typical values of $\sim 1~\rm{km \, s^{-1}}$ and therefore likely supported by turbulent pressure rather than thermal pressure alone. This turbulent support should lead to higher accretion rates in high-mass star formation \citep{McKee+2003,TanPPVI+2014}, where the accretion rate depends on $\sigma_{NT}$. Hence, for a massive core that is roughly virialized the accretion rate for high-mass star formation is much larger than the value given by equation~\ref{eqn:mdot} and is instead given by \cite[e.g.,][]{KrumholzBook2015, McKee+2003}
\begin{equation}
\dot M \approx \frac{\sigma_{NT}^3}{G} = 10^{-4} \msun {\rm yr}\e. 
\label{eqn:mdotmassive}
\end{equation}
The higher accretion rates inherent in high-mass star formation allow for faster formation timescales and larger ram pressures associated with the accretion flow which may counteract the pressures associated with stellar feedback as the stars contract to the main-sequence.

While numerical hydrodynamic simulations, which we describe next, have enabled models with increasing degrees of physical complexity, observations often remain limited to measurements of $T$, $\rho$, $R$, $B$ and $\sigma$. Thus the expressions above, which depend only on these fiducial parameters, remain useful benchmarks of the fundamental physical processes in star formation.

\subsection{Numerical Modelling of Star Formation with Hydrodynamic Simulations}

Star formation simulations typically adopt one of two main approaches to model gas dynamics: {\it grid}- or {\it particle}- based methods. We summarize these here and refer the reader to the review by \citet{Teyssier2019a} for a more detailed description of numerical methods used in simulating star formation. 

{\it Grid}-based methods discretize the partial differential equations of hydrodynamics and subdivide the computational domain into individual volume elements centered on node points distributed according to a grid or unstructured mesh. Grid approaches may adopt a fixed volume or ``cell" size for the entire domain (fixed-grid approach) or may adaptively change the cell size to enable finer resolution on selected small scales, i.e., adaptive mesh refinement (AMR) approaches. The advantage of AMR is that the user has flexibility to refine on specific quantities of interest such as density and velocity gradients.  Therefore, AMR methods are ideal for star-formation simulations, which span several orders of magnitude in spatial scale (i.e., from pc down to sub-AU scale).  {\it Grid}-based methods are uniquely suited to modelling high-mach number, magnetized flows, since they enable high-accuracy shock-capturing (small diffusivity) and robust treatment of magnetic wave propagation \citep{Teyssier2019a}. \add{However, AMR methods can produce round off errors at grid interfaces that can lead to advection errors, angular momentum conservation errors, and excessive diffusion \citep{Berger1989}}. A number of AMR grid-based codes are in use in star formation studies including, {\sc zeus-mp}, {\sc flash}, {\sc enzo}, {\sc ramses}, \add{{\sc pluto}}, {\sc orion}, and {\sc athena} \add{\citep{Flash00, Ramses02, Zeus2006, Athena08, Mignone2012, Li2012a, Enzo2019}}. 

Alternatively, star-formation calculations employ {\it particle}-based methods, like smoothed particle hydrodynamics (SPH), to model the hydrodynamic evolution of a fluid by discretizing the gas into a set of particles with mass and momentum. The evolution of the system is described by the motions of the large ensemble of interacting particles. The bulk hydrodynamic properties are then obtained by averaging over the particle distribution. SPH methods are ideal for problems modelling gravitational collapse, since particle behavior naturally provides adaptive resolution \add{ and modern SPH codes typically do not have conservation errors like AMR}. However, one of the disadvantages of SPH as compared to AMR, is that SPH codes lack the ability to sharply resolve shocks and therefore use an artificial viscosity to improve their shock capturing abilities \citep{Teyssier2019a}. This effect makes SPH codes less ideal than {\it grid}-based codes for simulating high-mach number flows and fluid instabilities that are common in star formation \citep{Tasker2008a}. Popular, public SPH codes include {\sc gadget}, {\sc gasoline}, and {\sc phantom}  \citep{Springel2005,Wadsley+2004,Price+2018}. 

Lagrangian ``moving-mesh" hydrodynamic codes provide an attractive alternative to SPH and AMR codes, combining the strengths of both, including high-numerical accuracy for shocks, low numerical viscosity and dynamic adaptivity. \add{However, like AMR some moving-mesh codes can lead to angular momentum conservation errors \citep{Hopkins2015}.} Two recently developed, publicly available codes include {\sc gizmo}, a hybrid moving-mesh, SPH code  \citep{Hopkins2015}; and {\sc arepo} an unstructured moving-mesh code \citep{Springel2010}. Both include formalisms for treating magnetic fields, stellar feedback, and dark matter, so they are also widely used for cosmological applications.

Although AMR and SPH methods allow spatial or mass refinement in star formation simulations across a significant spatial scale it is currently computationally \add{challenging and expensive} to follow the gravitational collapse of the ISM on the size scales of clouds and cores down to stellar size scales\add{, and follow the protostellar evolution for a significant amount of time}. In light of \add{these limitations}, sub-grid models are used to model the formation and evolution of (proto)stars with accreting Lagrangian sink particles \add{\citep{Bate1995, Krumholz2004a}} and their subsequent stellar feedback. For example, most star formation simulations follow the collapse of star forming regions in simulations by refining on the Jeans length \add{in AMR codes} given by
\begin{equation}
    \lambda_{\rm J} = \frac{c_s}{\sqrt{4\pi G\rho}}
\end{equation}
where $\rho$ is the gas density and $c_s$ is the sound speed, which describes the interaction between pressure and self-gravity where length scales less than $\lambda_{\rm J}$ are prone to gravitational collapse. To model this collapse, AMR simulations that model star formation usually apply the Truelove criterion \citep{Truelove1997} which resolves the local Jeans length with 4 cells or more (i.e., such that $\lambda_{\rm J} < 4 \delta x_{l}$ where $\delta x_{l}$ is the cell size on AMR level $l$). Sink particles that can accrete nearby gas are then placed in cells when the finest level exceeds the Truelove criterion on the finest level. We also note that this criterion may also take into account magnetic pressure if magnetic fields are present \citep{LeeA+2014}. The influence of magnetic pressure increases the Jeans length, thereby potentially suppressing star formation. \add{In contrast, in SPH simulations where the fluid is modeled as Lagrangian mass particles that are smoothed over  a weighting kernel, the local Jeans mass must be resolved with a minimum of a $2 N_{\rm Neigh}$ particles where  $N_{\rm Neigh}$ is the number of particles in the SPH kernel to properly model fragmentation \citep{Bate1997}. Sink particles can then be placed once this critical density is reached.}

These particles are then modeled with a protostellar prescription describing their evolution while they accrete from the surrounding gas. As they evolve and grow in mass they then inject momentum and energy to nearby cells via sub-grid prescriptions that model stellar feedback like radiation, collimated outflows, and stellar winds. These sub-resolution models reduce the computational time of star formation simulations while \add{including observationally motivated physical processes that are present during star formation.}

Numerical calculations require as inputs the initial values for each of the fundamental variables in the problems. Bulk properties such as temperature, mean density, mean magnetic field and $\sigma(R)$ can be drawn from observations; however, the exact 3D starting distributions of gas densities, velocities and magnetic field is less certain, since exactly what constitutes the initial conditions of star-formation, particularly on sub-pc scales, is not a well-posed problem.

To minimize computational expense calculations often begin by either modeling an full cloud by assuming a $\sim$ few-10 pc sphere, modeling a piece of a molecular cloud by adopting periodic boundary conditions that encompass a few pc, or beginning with individual cores of size scale $\sim0.1$ pc and adopting analytic conditions as described in Section~\ref{sec:corecol}. Since this chapter is devoted largely to the formation of individual star systems, we focus on the latter calculations, which often follow an isolated, pressure confined sphere of gas and span sub-pc to AU scales in Section~\ref{sec:SFtheory}. One key advantage of focusing on the core scale is that it enables the consideration of a broader range of physics. Here we focus on calculations of low- and high-mass star formation that includes some combination of magnetic fields, turbulence, radiative transfer and stellar feedback. We note that including all physical effects, achieving sub-AU resolution, and evolving the equations over the full formation timescale remains beyond computational resources (and human patience). 

\subsection{The Pre-cursors to Stars: Formation of the First and Second Hydrostatic Cores}

\label{sec:fhsc}

The early onset of star formation involves the gravitational collapse of a pre-stellar core \add{or collection of dense material, possibly collected by colliding shocks, that becomes gravitationally unstable} to form a hydrostatic object known as the ``first core" that is supported by its own internal pressure. Several numerical studies using both grid-based \add{\citep[][]{Bodenheimer1968, Winkler1980a, Winkler1980b, Stahler1980a, Stahler1980b, Stahler1981, Masunaga1998, Masunaga2000, Tomida2010b, Commercon2011b, Vaytet2012, Tomida2013, Vaytet2013, Vaytet2017, Vaytet2018, Bhandare2018}} and SPH methods \add{\citep{Whitehouse2006, Stamatellos2007, Bate2014, Tsukamoto2015, Wurster2018b}} have deduced that star formation occurs via a two-step process of the formation of first and second quasi-hydrostatic Larson cores that form from the gravitational collapse of pre-stellar cores \citep{Larson1969}. As the system evolves further, the conservation of angular momentum leads to the formation of a circumstellar disk around the central protostar, which can eventually host companion stars and/or planet(s). In order to strengthen our understanding of how stars form, it is crucial to perform robust and detailed self-consistent studies of the transition of a pre-stellar core to a second hydrostatic core that can eventually become a star.  

\begin{figure}[tp]
	\centering
	\includegraphics[width=0.8\linewidth]{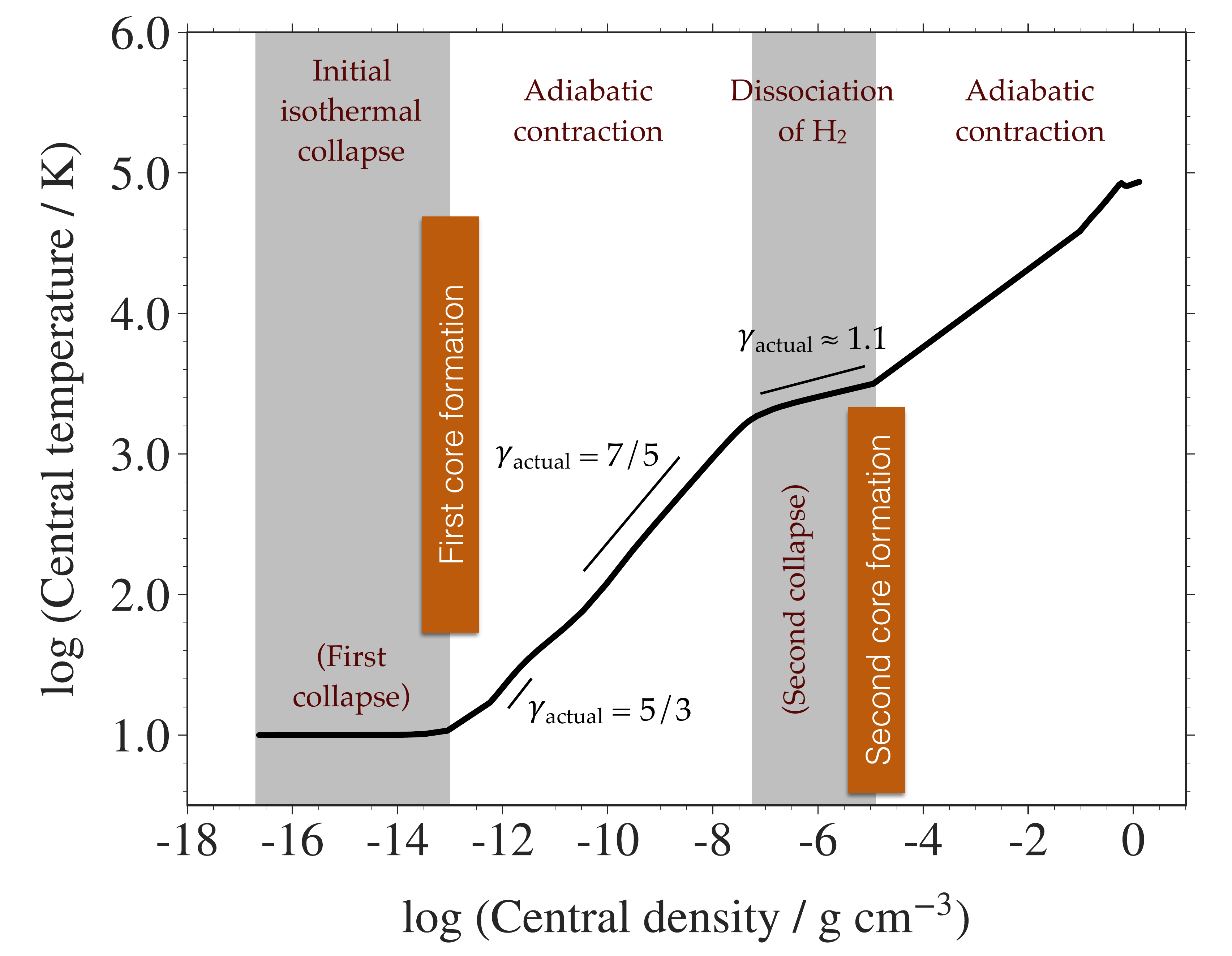}
	\caption{Thermal evolution showing the first and second collapse phase\add{s} for a 1 $M_{\odot}$ cloud core with a fixed outer radius of 3000 au. The change in adiabatic index $\mathrm{\gamma_{actual}}$ indicates the importance of using a realistic gas EOS. Figure reproduced from \citet{Bhandare2018}. Reproduced with permission \copyright ESO.}
	\label{fig:thermalevolution}
\end{figure}

The process of formation of the first and second quasi-hydrostatic Larson cores is indicated in Figure~\ref{fig:thermalevolution} and can be summarized as follows. Due to efficient cooling via thermal emission from dust grains and line emission from molecular gas, the initially optically thin, isothermal cloud core collapses under its own gravity. Gravitational collapse in molecular cloud cores can be triggered either by the support of ambipolar diffusion of magnetic fields \citep[e.g.][]{Shu1987, Mouschovias1991}, by the dissipation of turbulence in prestellar cores \citep[e.g.][]{Nakano1998}, or an external shock wave crossing a previously stable cloud \citep{Masunaga2000} can also lead to the collapse. 

As the density increases during the initial isothermal phase, the optical depth becomes greater than unity and radiative cooling becomes inefficient. With time, as the cloud compresses, temperature in the dense parts begins to rise. This leads to the first adiabatic collapse phase, which is followed by the formation of the first hydrostatic Larson core. At this stage the gas behaves as monatomic and the first core eventually contracts adiabatically with an adiabatic index $\gamma_{\mathrm{actual}}$~$\approx$~5/3, where $\gamma_{\mathrm{actual}}$ is the change in the slope of the temperature evolution with density. As the temperature increases, the rotational and vibrational degrees of freedom for the $\mathrm{H_2}$ molecules start being excited as the cloud transitions from being effectively monatomic to diatomic. During this phase, the adiabatic index changes to $\approx$ 7/5. As an example, formation of the first hydrostatic core takes roughly $\mathrm{10^4}$ years during the collapse of a 1 $\mathrm{M_{\odot}}$ cloud core with a size of 3000 au and an initial temperature of 10 K. The first collapse halts when the central density is of the order of $\mathrm{10^{-13}~g~cm^{-3}}$. On average, the size of the first hydrostatic core is roughly a few au. 

Once the central temperature reaches $\sim$ 2000~K, $\mathrm{H_2}$ molecules begin to dissociate. $\mathrm{H_2}$ dissociation is a strongly endothermic process, which allows gravity to dominate over pressure and initiates the second collapse phase. At typical central densities of $\mathrm{10^{-8}~g~cm^{-3}}$, 
	\add{the core undergoes second collapse.} The second hydrostatic core is formed after most of the $\mathrm{H_2}$ is dissociated and eventually undergoes a phase of adiabatic contraction. The formation phase of the second hydrostatic core is comparatively much faster and lasts only for a few hundred years. The second hydrostatic core forming within the first hydrostatic core has an initial size on sub-au scales. An increase in thermal pressure halts the collapse, while the second hydrostatic core continues to accrete material from its surrounding envelope and can grow in mass. A star is born once the core reaches ignition temperatures (T $\geq 10^6$ K) for nuclear hydrogen burning. 

Evolving the second core until the protostellar phase has been challenging mostly due to resolution (i.e., time step) limitations. Thus, the non-homologous collapse phases of first and second core formation have been extensively investigated using one-dimensional studies. The focus of modern collapse studies has been on the microphysics of these hydrostatic cores by including a realistic gas equation of state (to account for the effects of $\mathrm{H_2}$ dissociation, ionization of atomic hydrogen and helium, and molecular rotations and vibrations), dust and gas opacities, as well as an accurate treatment of the radiation transport \citep[see the recent review by][for various numerical methods]{Teyssier2019a}. The \add{1D} collapse simulations by \citet{Vaytet2012, Vaytet2013} suggest that multi-group (i.e., frequency dependent) radiative transfer would prove to be important in the much later stages during the long-term evolution of the second core. \add{The 1D numerical} studies by \citet{Masunaga2000} have been the pioneers of these self-consistent simulations and the only ones, so far, to evolve the second hydrostatic core until the end of the main accretion phase. 

Recent \add{1D simulations} by \citet{Vaytet2017} span a wide range of initial molecular cloud core properties such as cloud size, initial temperature, mass, and density distribution (uniform vs Bonnor-Ebert \citep[][ see Section~\ref{sec:corecol}]{Bonnor1956}). These protostellar collapse models focus on the low-mass regime (i.e., for initial cloud core masses up to 8~$M_\odot$) \add{and they find that the properties of the first and second cores are mostly insensitive to the initial cloud properties}. Following the same principle, \citet{Bhandare2018} expanded these collapse studies to cover the parameter space in the intermediate- and high-mass regimes using initial cloud masses from 0.5~$M_\odot$ to 100~$M_\odot$. Both of these studies 
establish\add{ed} quantitative estimates for the properties of the first and second cores. As a strong distinction between \add{the} low- and high-mass regimes, the first hydrostatic cores are seen to be non-existent in the high-mass regime due to high accretion rates \citep{Bhandare2018}. This provides a useful constraint for observational efforts in detecting first hydrostatic core candidates. 

\add{The 1D} studies \add{mentioned above} provide a lower bound on more realistic hydrostatic core properties, especially lifetime estimates, derived by accounting for the effects of initial cloud rotation, turbulence, and magnetic fields. The kinetic (rotational and/or turbulent) and magnetic support in two- and three-dimensional simulations can slow down the collapse \citep{Tomida2013}. Additionally, these multidimensional simulations prove to be valuable in order to trace the formation and evolution of circumstellar disks formed around young stars due to conservation of angular momentum of the infalling material. Some numerical studies have found that the first hydrostatic core evolves into a disk even before the onset of the second core formation \citep{Bate1998, Bate2010, Bate2011, Machida2010, Machida2014, Tomida2015, Wurster2018b, WursterBate2018}. On the contrary, other studies have found that the disk is formed only during or after the formation of the second hydrostatic core \citep{Dapp2010, Machida2011, Dapp2012, Tomida2013, Machida2014, Tomida2015, Tsukamoto2015, WursterBate2018, Vaytet2018}. This discrepancy has a strong dependence on the initial conditions, the included physics, and the evolution of the collapsing cloud as described in the recent review by \cite{WursterLi2018}.

The effects due to self-gravity, a realistic gas equation of state, radiative transfer and non-ideal (including ohmic and ambipolar) MHD on the formation of the first and second hydrostatic cores is captured in more recent studies using grid-based \add{\citep{Tomida2013, Tomida2015, Vaytet2018}} and SPH \add{\citep{Tsukamoto2015, Wurster2018}} codes. The first two thousand years of pre- to protostellar evolution are recently traced using 3D resistive MHD simulations using a barotropic equation of state \citep{Machida2019}. Currently, a parameter scan using different initial conditions or long-term calculations for isolated 3D radiation-MHD collapse simulations, which resolve the first and second hydrostatic cores is still not possible owing to time step restrictions\add{, which is a common feature in all simulations discussed thus far when modeling second core formation}. One possible solution for investigating the long-term evolution of the protostellar core is to replace the second core with a sink particle using sub-grid models. Future work in cloud collapse calculations to properly capture first and second hydrostatic core formation should include additional effects due to chemistry and a multi-fluid approach, which would also account for the dynamics of decoupled dust grains. This would play an important role in determining the cooling efficiency and opacities as well as aid the resistivity calculations for non-ideal MHD.

\subsection{Hydrodynamic Simulations of Low- and High-Mass Star Formation}

There are four main questions that multi-physics hydrodynamical simulations of  star formation aim to address: What are the duration and characteristics of accretion? What is the star formation efficiency of the dense gas? How does angular momentum transport occur? What is the role of magnetic fields on different scales? Numerical simulations are an indispensable tool for addressing these questions, which require the consideration of multi-scale, non-linear physics acting in concert.

{\it The Role of Magnetic Fields:} In the 1970s, before the turbulent nature of molecular clouds was recognized, star formation was thought to be regulated by magnetic fields \citep{MouschoviasSpitzer1976,Mouschovias1976}. In the limit of ideal MHD, flux conservaton implies that initially magnetically supported (``sub-critical", $\mu_\phi < 1$) cores would never go on to collapse. This problem is resolved by non-ideal magnetic effects, such as ambipolar diffusion, which allow magnetic fields to diffuse out of the cores. Although direct observations of magnetic field strengths remain challenging, Zeeman observations suggest that dense cores tend to be mildly supercritcal with $\mu_\phi \sim 2$ \citep{Crutcher2012}. Synthetic Zeeman observations of numerical simulations with strong magnetic fields on the cloud/clump scales show good agreement with these results \citep{LiMcKee+2015}. \add{Additionally, magnetic pressure, which opposes gravity, reduces fragmentation in high-mass clouds and cores, potentially aiding high-mass star formation \citep[e.g.,][]{Hennebelle2011, Myers+2013}}.


\add{The complex structures observed in \add{high-resolution} dust polarization \add{studies} lend support for the sub-dominance \add{\textit{or}} dominance of magnetic fields within cores and protostellar envelopes. For example, the small scale structure on scales of $\sim 0.01$\,pc \add{observed in Ser-emb 8, a low-mass Class 0 protostar in the Serpens Main star-forming region,} can be reproduced by simulations with relatively weak initial magnetic fields ($\beta > 0.25$ where $\beta$ is the plasma parameter and is the ratio of thermal pressure to magnetic pressure) \citep{Hull+2017}. In contrast, \citet{Maury2018}  find that the Class 0 protostar B335 envelope, at 50-1000 au scales, has an ordered magnetic field topology with a transition from a large-scale poloidal magnetic field, in the outflow direction, to strongly pinched in the equatorial direction suggesting that the field lines are being dragged in during collapse. Indeed, detailed zoom-in numerical simulations by \cite{Hennebelle2018}, which measure the properties of a large sample of prestellar cores, have demonstrated that a large scale weak magnetic field within the cloud will lead to a diversity of magnetic field strengths in prestellar cores, with cores having typical $\mu_{\phi}$  values that range from 0.3-3 (i.e., from highly sub-critical to super-critical values).} \add{Observations  agree with this magnetic field diversity in prestellar cores.} \cite{HullZhang19} compiled a comprehensive review on the interferometric observations of low- and high-mass star formation that describes the magnetic field strengths and morphologies for a sample of low-mass cores and high-mass star forming clumps. By combining the literature of these observations they found that, on the size scales of 0.1-1 pc, 28\% of the low-mass star forming cores and 21\% of the high-mass star forming clumps with known field morphologies exhibit pinched hour-glass magnetic field morphologies. \add{These results suggest that the scenario of magnetically dominant core collapse may not be the predominant mode of low- or high-mass star formation, but is dominant in a non-negligible fraction of sources.}

Magnetic fields also influence disk formation and are likely responsible for producing the collimated jets and entrained outflows that are ubiquitous in low- and high-mass star formation (e.g., see Section~\ref{sec:obs}). These outflows are either launched by magneto-centrifugal acceleration at the interface of the stellar magnetosphere with the disk \citep[X-wind model,][]{Shu1995}, by magneto-centrifugal processes from the centrifugally supported part of the disk \add{\citep{Blandford1982, Pelletier1992}}, or by magnetic pressure alone \citep{LyndenBell2003}\add{, which we discuss in more detail next}. 


{\it Angular Momentum Transfer:} One of the classic problems in star formation concerns the observation that the angular momenta of clouds and cores is orders of magnitude larger than that of stars \citep{Bodenheimer1995}. Since angular momentum is conserved by nature, some process or processes must drive the transport of angular momentum and facilitate collapse. Numerical simulations suggest that magnetic and gravitational torques can provide the necessary lever arm to move angular momentum from smaller to larger scales \citep{Armitage2011, Lin2011, Rosen2012, KratterLodato2016}. Magnetic \add{braking}, in which the coupling between the gas and field allows high-angular material to follow field lines outwards, acts on both core and disk scales. In the limit of ideal MHD for strong magnetic fields ($\mu_{\rm \Phi}<10$), magnetic braking removes angular momentum so efficiently that no accretion disk forms \add{whereas disk braking will be reduced for weaker magnetic fields and lead to accretion disks with reduced disk sizes as compared to those form from un-magnetized gas} \citep{Commercon2011, Seifried2011, LiPPVI+2014}. \add{ Turbulence-induced misalignment (i.e., misalignment between the angular momentum axis and magnetic field)  and magnetic diffusivity can significantly reduce magnetic braking, leading to disk masses and sizes that are reduced as compared to when magnetic fields are absent  \citep{Joos2012, Joos2013, Seifried2013, Gray2018}.} Additionally, non-ideal MHD effects, including the Hall effect, Ohmic resistivity, and ambipolar diffusion help reduce the efficiency of angular momentum transport, thus enabling disk formation \add{\citep{Wurster+2016,Zhao+2016, Kolligan2018, WursterLi2018}}. Turbulence and misalignment of the magnetic field and angular momentum vectors also mitigate the angular momentum problem \add{\citep{PriceBate2007a,HennebelleCiardi2009,Krumholz+2013, Myers+2013, WursterLi2018}}. Once a disk forms, gravitational instability and the magneto-rotational instability (MRI) act to transport angular momentum in the outer and inner disk, respectively, thus allowing accretion to continue. \add{For more evolved disks that surround pre-main sequence stars (e.g., magnetically dead disks), non-ideal effects can reduce angular momentum removal by MRI and instead drive a disk wind to remove angular momentum to aid accretion \citep{Bai2013, Gressel2015}}.


Turbulence and core angular momentum is inextricably linked.  Turbulence by nature contributes angular momentum and rotational energy; the largest scale turbulent mode can explain the magnitude of observed core velocity gradients \citep{BurkertBodenheimer2000,ChenOstriker2018}.
Numerical simulations following the collapse of low-mass, turbulent magnetized cores show that their angular momentum goes as $j = L/M \propto r^{3/2}$ \citep{ChenOstriker2018}. The presence of turbulence supplies angular momentum and can explain the magnitude of velocity gradients observed on core scales \citep{BurkertBodenheimer2000,ChenOstriker2018}. The time and spatially changing angular momentum of turbulent gas produces outflow direction variation \citep{Lee+2017} and contributes to misaligned outflows and spins characteristic of wide binary systems \citep{Offner+2016}.

{\it Accretion Timescale:} From start to finish the stages of star formation comprise of core collapse, accretion, and arrival on the main sequence. For low-mass star formation this sequence of events spans millions of years owing to their long formation time scales ($\sim 1$ Myr) and longer contraction Kelvin-Helmholtz timescales (the time required for a star to radiate away its gravitational binding energy and contract to the main sequence). In contrast, \add{t}he formation time scales for high-mass stars are much shorter and span $\sim10,000s-100,000$ yr due to their short Kelvin-Helmholtz timescales. Therefore, high-mass stars attain their main sequence luminosities while they are still  actively accreting. Here we focus on the stage that comprises the main collapse and accretion phase. However, the mapping between class and true evolutionary state is confused by projection effects, in which edge-on sources appear younger, and accretion variation \citep{Robitaille+2006, Offner+2012b,DunhamVorobyov2012}. 

When full physical information is available it is possible to define more physically motivated stages that are independent of the protostellar bolometric temperature and luminosity. The majority of accretion occurs during ``Class 0," which is defined as when the protostellar mass is less than the envelope mass, $M_p < M_{\rm env}$, or alternatively, when at \add{least} half the initial envelope has been accreted or expelled, $M_{\rm env}> 0.5 M_{\rm env,0}$ \citep{DunhamPPVI+2014}.  
Hydrodynamic simulations of low-mass core collapse indicate that the Stage 0 lifetime lasts $\sim$0.1-0.3\,Myr, where calculations with no turbulence and weaker magnetic fields collapse faster \citep{MachidaHosokawa2013,OffnerChaban2017}. This timescale is much longer than typical core free-fall times (Eq.~\ref{eqn:tff}) of dense cores, since magnetic fields and turbulence provide additional pressure support, retarding the rate of collapse and accretion.

The details of the initial core structure and turbulence also impact the collapse timescale. Turbulence with a larger ratio of compression ($\nabla \cdot v \neq 0$) to stirring ($\nabla \times v \neq 0$) facilitates core collapse and fragmentation \citep{Girichidis+2012}. Turbulence driven by protostellar outflows has a higher solenoidal to compressive ratio and thus tends to slow rather than promote collapse within dense cores \citep{Hansen+2012,OffnerArce2014,OffnerChaban2017}.  In the case of low-mass star formation, radiation has little effect on the rate of collapse \citep{PriceBate2007a,OffnerArce2014}, although it does reduce the incidence of fragmentation and raise the primary stellar mass (see Section~\ref{sec:lowmassfeedback}).

\begin{figure}
\begin{center}
\includegraphics[width=1\linewidth,trim={0 11cm 0 0},clip]{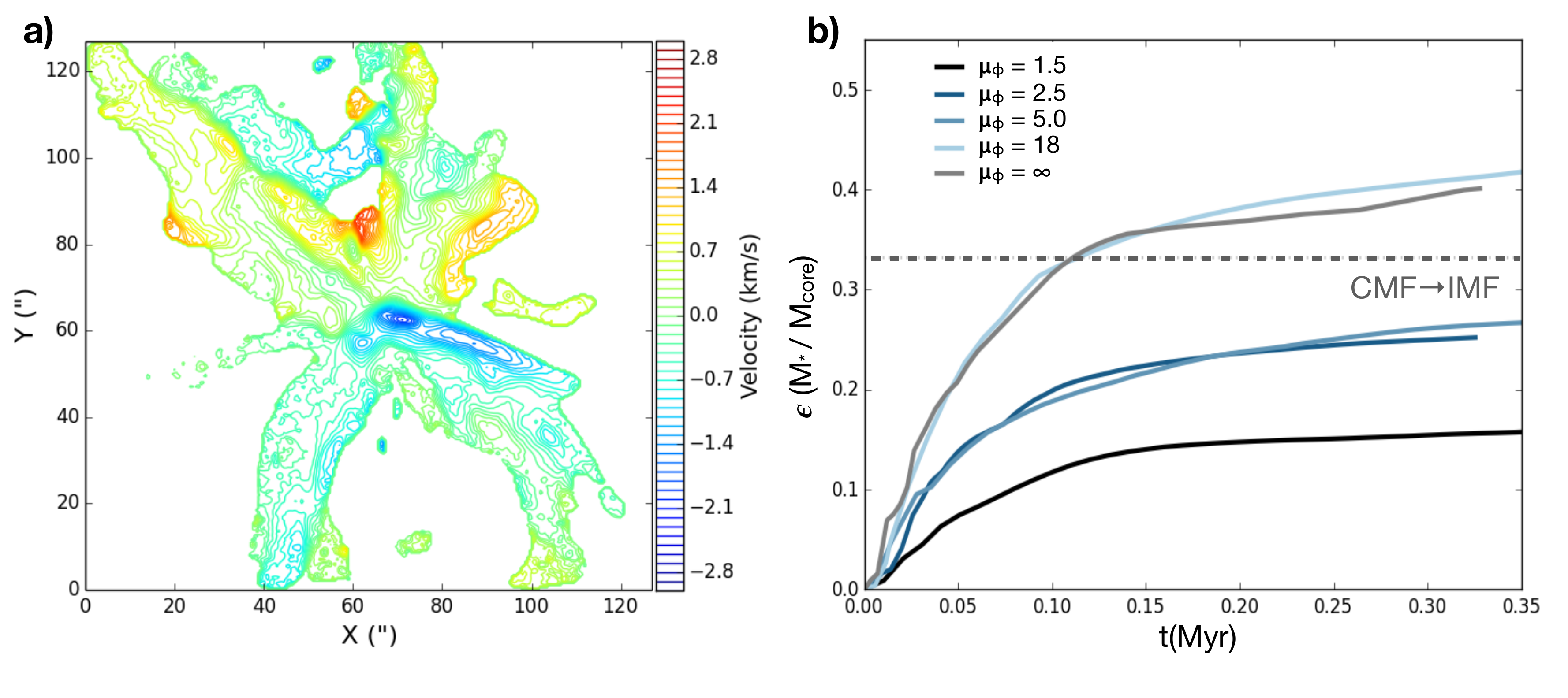}
\label{fig:lmt}
\caption{Left (a): Synthetic $^{13}$CO(1-0) ALMA observation of a simulated protostellar outflow \citep{Bradshaw+2015}. Right (b): Dense gas efficiency versus time for simulations of accreting low-mass protostars with different initial magnetic field strengths \citep{OffnerChaban2017}. The cores have initial masses of 4\,\msun and initial turbulence with $\sigma = 0.72$\,km s\e. The dotted line represents the efficiency to map the observed CMF peak to the IMF peak. Both panels \textcopyright AAS. Reproduced with permission. }
\label{lowmassoutflows}
\end{center}
\end{figure}

{\it Star Formation Efficiency:} The observed core mass function (CMF) strongly resembles the stellar IMF but is shifted to higher masses by a factor of $\sim $3 \citep{Alves07}. This suggestive similarity reinforces the idea that the efficiency of even dense gas ($n>10^4$\,g\,cm$^{-3}$) is relatively low. One explanation for the CMF/IMF offset is feedback, namely, protostellar outflows acting on $\sim0.1$\,pc core scales produce this inefficiency by entraining and expelling a large fraction of the core \citep{Hansen+2012,Federrath2015,Cunningham2011, Kuiper2016}. Numerical simulations of isolated core collapse provide the most straightforward way to parameterize the impact of outflows as a function of core properties.
Such simulations suggest outflows have a mass-loading factor of $\sim$3 \citep{Kuiper2016, OffnerChaban2017}, with efficiency declining with the magnetic field strength of the natal core as shown in Figure~\ref{lowmassoutflows}. The efficiency of entrainment may be affected by the outflow opening angle \citep{MachidaHosokawa2013,OffnerArce2014}. There is some evidence that outflows widen with time on average \citep{ArceSargent2006,Offner+2011,Kuiper2016}, but simulations also demonstrate that individual outflows evolve non-monotonically \citep{Offner+2011,MachidaHosokawa2013}. In the context of high-mass star formation, additional feedback from radiation pressure, photoionization, and stellar winds can also inhibit accretion and expel material from accreting high-mass stars \citep{Rosen2016, Kuiper2018, Rosen2019}.

The star formation efficiency may also be influenced by fragmentation and dynamics \citep{Holman+2013,OffnerPPVI+2014,Rosen2019}. The relatively high occurrence of binary and triple systems suggests that multiplicity does act to lower the core to star efficiency. However, radiative feedback and magnetic pressure support tend to reduce the formation of higher-order multiples, by reducing additional fragmentation \citep{PriceBate2007b,Offner+2009,Commercon2011,OffnerArce2014, Fontani2018, Rosen2019}. In the following Sections, we discuss the mechanisms and impact of feedback in low-mass and high-mass star formation, respectively.

\subsection{\add{Stellar} Feedback in Low-Mass Star Formation} 
\label{sec:lowmassfeedback}

Although the process of low-mass star formation is considerably less violent and energetic than high-mass star formation, stellar feedback nonetheless plays an important role in shaping the outcome. Radiative feedback and magnetically launched protostellar outflows are the dominant feedback processes for stars with masses $\lesssim 5$\,\msun.  Cosmic rays accelerated by protostellar accretion and outflow shocks may also play an important role locally in shaping the chemistry, temperatures and accretion disk evolution, if not the broader gas dynamics. We discuss the impact of each of these feedback processes below. 

\subsubsection{\add{Radiative Heating}}  

Low-mass protostars are relatively dim compared to their high-mass counterparts with luminosities spaning only $\sim 0.1-10^2$\,\lsun \citep{DunhamPPVI+2014}. However, these modest luminosities, which are produced primarily by protostellar accretion, are sufficient to heat gas 10s to 100s\,K within a few  hundred au of the source \citep{Offner+2009}. The elevated gas temperatures provide additional thermal support to the accretion disks, thereby significantly increasing disk stability and reducing the incidence of disk fragmentation \citep{Bate2009,Offner+2010,Bate2012}. This in turn reduces the number of brown dwarfs and low-mass stars formed, consequently, enabling good agreement with the observed stellar IMF \citep{Bate2009,Bate2012,Krumholz+2012}. Thus, radiative feedback, which is in some sense
self-regulating, is an essential ingredient to reproduce the surprisingly invariant stellar IMF that is observed \citep{Krumholz+2016,Guszejnov+2016,Cunningham+2018}. 

Due to its impact on small-scale fragmentation, radiative feedback also plays an important role in shaping stellar multiplicity. About half of all solar-type stars reside in multiple systems \citep{Raghavan+2010}. Without the influence of local radiative heating, low-mass binary star systems are created primarily by disk fragmentation and the subsequent dynamical evolution of small n-body systems \citep{Bate2009a,Lomax+2015}. Under this scenario, nearly all low-mass stars are born in multiple systems. 

In the absence of frequent disk fragmentation, a large number of stellar multiples are instead formed by fragmentation at wide separations (``turbulent core fragmentation"), which then migrate on $\sim 0.1$\,Myr timescales to sub-100\,au separations \citep{Offner+2010,Offner+2016,Kuffmeier+2019,Lee+2019}. If the spins of protostellar pairs remain misaligned even after migration, stellar spin orientation may serve as a signpost to distinguish between different binary formation mechanisms long after the natal gas is dispersed \citep{Offner+2016}.
Radiative feedback naturally explains the difference in multiplicity between low- and high-mass stars, which likely experience a high degree of disk fragmentation \citep{KratterLodato2016}. However, the relative incidence of disk versus core fragmentation also depends on the initial conditions of star formation and other physical properties including magnetic field strength, and thus, remains debated \add{since magnetic fields, in addition to radiative heating also suppresses fragmentation \citep{Price2009a, Commercon2010a, Myers+2013, Cunningham+2018}.}

Observations suggest that radiative feedback is strongly time-variable, cycling between short periods of high-luminosity and longer periods of low luminosity \citep{Audard+2014}. However, the frequency of luminosity bursts, which depends on both the details of protostellar evolution and accretion microphysics, is not well constrained. Several theoretical models with different degrees of accretion variation successfully reproduce the protostellar luminosity distribution observed in low-mass star-forming regions \citep{DunhamPPVI+2014}. Gas chemistry provides another avenue of constraining the magnitude of radiative feedback. If the gas near protostars, $r < 10^3$\,au, was significantly warmer in the past, e.g., due to an accretion burst, then molecules that are typically frozen out at high densities and cold temperatures, such as CO, will be present in the gas phase \citep{Jorgensen+2013,Vorobyov+2013}. Astrochemistry models that combine hydrodynamics and gas-grain chemical networks provide a promising means to constrain the time variation of radiative feedback. Increasingly high-resolution studies of disk properties and complex core chemistry will help discriminate between radiative feedback models.

\subsubsection{Protostellar Outflows} 
\label{sec:outflows}
Magnetically launched, collimated protostellar outflows play a significant role in setting the star formation efficiency and gas dynamics of dense low-mass cores. Simulations of forming low-mass stars suggest that the momentum injected by outflows entrain and expel 30-50\% of the gas in dense cores \citep{MachidaHosokawa2013, OffnerArce2014, OffnerChaban2017}. The efficiency of entrainment depends on the initial turbulence and magnetic field strength, with stronger fields producing lower star-formation efficiencies \citep{OffnerChaban2017}. The effective mass loading factor is $\sim 3$ and appears largely invariant to core properties \citep{OffnerChaban2017}. Outflows, by clearing local dense gas, provide the only mechanism to shut off protostellar accretion in low-mass star formation, thereby helping to set the peak mass of the stellar IMF \citep{Hansen+2012,Cunningham+2018}. 

Protostellar outflows also inject significant momentum into their surroundings from core to cloud scales, driving turbulence and further reducing star formation efficiency globally \citep{Hansen+2012,Federrath2015,Cunningham+2018}. Numerical simulations demonstrate that the sustenance of turbulence by protostellar outflows may play a critical role in slowing global gravitational collapse and increasing cluster formation timescales for low-mass star clusters on size scales of $\sim 1$\,pc \citep{MatznerMcKee2000,NakamuraLi2007,Wang+2010}. Outflows may further drive turbulence by exciting magnetic waves, which propagate well-beyond the immediate outflow interaction region \citep{OffnerChaban2017,OffnerLiu2018}.

Protostellar outflows provide insights into the evolution of angular momentum of the star-disk system. Outflows and disks in wide multiple systems are frequently observed to be misaligned \citep{Williams+2014,Lee+2016}, 
which can be explained if the angular momentum of the initial protostars is not correlated \citep{Offner+2016} or if outflow orientation changes over time due to turbulence on small scales \citep{Fielding+2015,Lee+2017}. 

\subsubsection{Cosmic-Ray Feedback}

Recent observations and theoretical models suggest that magnetized protostellar accretion and outflow shocks may be sites of particle acceleration \citep{Padovani+2016}. Cosmic rays (CRs) accelerated by the star formation process via Fermi acceleration are relatively low energy ($<100$\,GeV) and thus dynamically unimportant on core scales \citep{GachesOffner2018}. However, this CR feedback may drive important chemical changes, which in turn affect gas heating and cooling \citep{GachesOffner2019}, disk formation \citep{Padovani+2014}, and disk stability \citep{Offner+2019}. 

\subsection{\add{Stellar} Feedback in High-Mass Star Formation}
High-mass star formation is influenced by many additional mechanisms of stellar feedback as compared to low-mass star formation because they contract quickly and attain their main-sequence luminosities while they are actively accreting. The formation time scale can be estimated by their Kelvin-Helmholtz timescale, $t_{\rm KH}=GM^2/RL$ where $M$, $R$, and $L$ are the final stellar mass, radius, and luminosity, respectively. For example, the formation time scale for 10 $M_{\rm \odot}$ and 50 $M_{\rm \odot}$ stars are $\sim$140 kyr and $\sim$20 kyr, respectively. Therefore, in addition to stellar outflows they also feedback on their surroundings via radiation pressure on dust, photoionization, and fast, isotropic stellar winds that are radiaitively driven from their surface. All of these feedback mechanisms have important implications for how material is delivered to the star and if feedback sets the upper mass limit of the IMF.

\subsubsection{\add{Radiative Heating and Pressure}}
Most theoretical attention in high-mass star formation studies has focused on radiation pressure on dust, which is the primary absorber of stellar radiation owing to its large opacity in dense star-forming environments \citep{Draine2003}. Early theoretical work suggested that high-mass stars with masses $\gtrsim 20 \, \msun$ should not form because radiation pressure will overturn the accretion flow \citep[e.g.,][]{Larson71a, Yorke77, Wolfire87}, contradictory to \add{modern} observations of \add{the IMF of} high-mass stars \citep[e.g.,][]{Schneider2018a}. For spherical (isotropic) accretion this effect can be parameterized by the Eddington ratio, $f_{\rm edd} = f_{\rm rad}/f_{\rm grav}$, which describes the relative importance of the radiative force ($f_{\rm rad}$) and the gravitational force ($f_{\rm grav}$) and is given by
\begin{equation}
\label{eqn:feddh}
f_{\rm edd} = 7.7 \times 10^{-5} \left( 1 + f_{\rm trap} \right)  \left( \frac{L_{\rm \star}}{M_{\rm \star}} \right)_{\rm \odot} \left( \frac{\Sigma}{1 \; \rm{g \; cm^{-2}}} \right)^{-1}
\end{equation}
where $\Sigma$ is the surface density of the dusty infalling material and $\left(L_{\rm \star}/M_{\rm \star}\right)_{\rm \odot}$ is the stellar light-to-mass ratio in solar units. The factor $\left( 1 + f_{\rm trap} \right)$ included in $f_{\rm rad}$ denotes the combined contribution from the direct radiation pressure associated with the first absorption of the stellar radiation field and the reprocessed thermal, diffuse radiation pressure associated with the re-emission by interstellar dust (parameterized by $f_{\rm trap}$), respectively. 

Given the importance of radiation pressure in high-mass star formation, it is crucial to include radiative transfer in \add{numerical} simulations. Early studies employed a gray (frequency averaged) flux limited diffusion (FLD) method, which is an approximation to the radiative transfer equation, to model the thermal dust-reprocessed radiation field inherent to the dusty ISM and a sub-grid model, in which \add{the protostar's radiative} energy is deposited \add{in a small volume} near the source, to model radiation from the accreting protostars \citep{Yorke2002, Krumholz2009a, Commercon2011}. However, \add{using a sub-grid prescription to model} stellar radiation greatly underestimates the radiation pressure. In light of this limitation, more recent simulations have employed a hybrid radiative transfer method that employed an FLD approximation to treat radiation pressure from the ISM and a \add{multi-frequency} ray tracing method \add{or frequency averaged (gray) M1 moment-based method} to properly treat the radiation field from accreting protostars \citep{Kuiper2010, Klassen2016, Rosen2016, Rosen2017, Mignon-Risse2020}.  

\begin{figure}
\begin{center}
\includegraphics[width=1\textwidth]{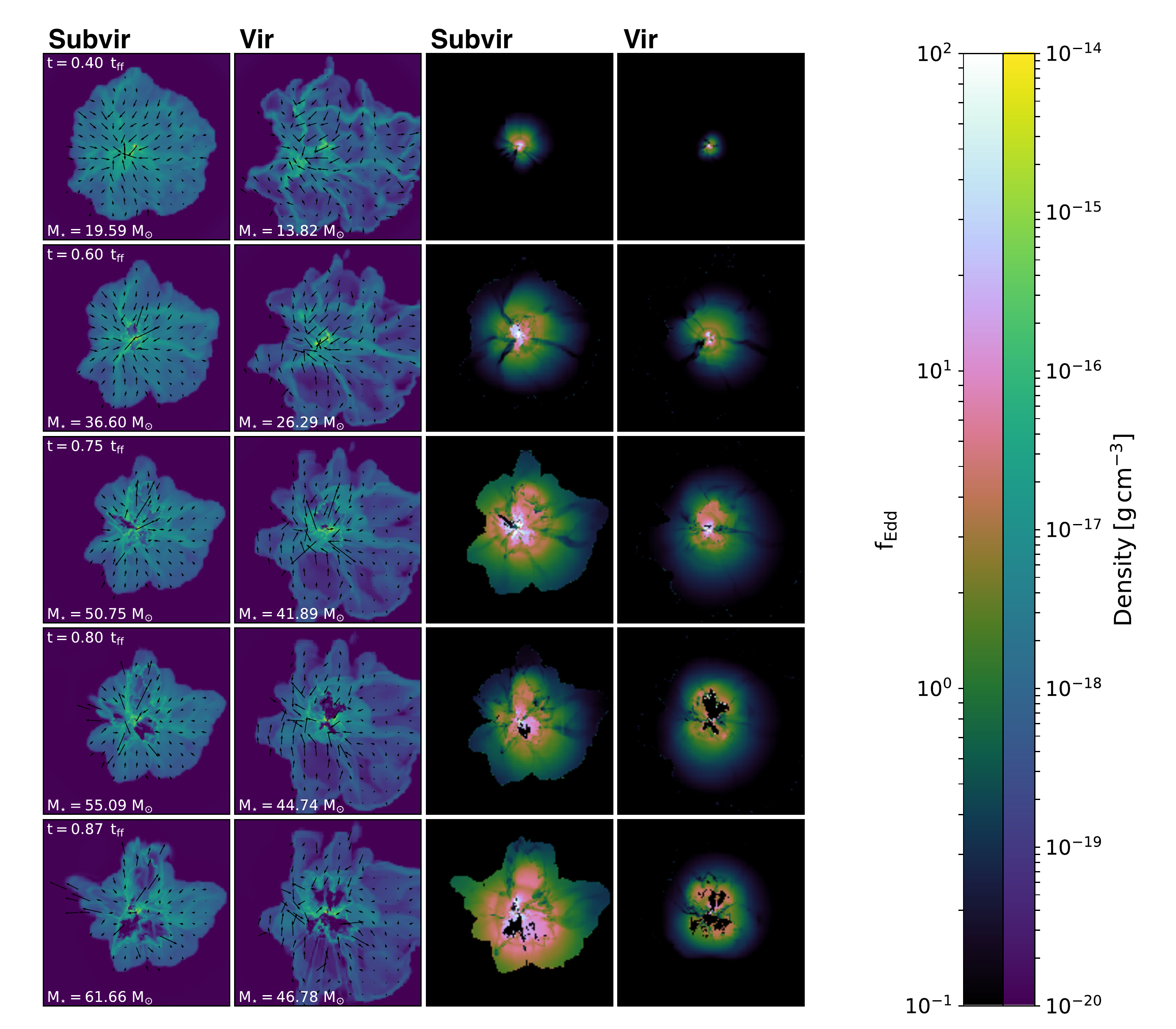}
\label{fig:hmt}
\caption{
 Snapshots of the density  and $f_{Edd}$ slices collapsing, turbulent 150 \msun prestellar core into a massive stellar system for a highly subvirialized core  ($\alpha_{\rm vir} = 0.14$; far left column (density), center right column ($f_{\rm Edd}$)) and roughly virialized core ($\alpha_{\rm vir} = 1.1$; center left column (density), rightmost column $f_{\rm Edd}$)) with velocity vectors overplotted. The time (in units of the core freefall time $t_{\rm ff}=42.7 kyr$) of the simulation and mass of the most massive star are given in the upper left corner of the far left panels and the lower left corner of the panels in the two left columns,
respectively. Figure taken from \citet{Rosen2019} \textcopyright AAS. Reproduced with permission.}
\end{center}
\end{figure}

With the use of these methods, numerous multi-dimensional radiation hydrodynamics (RHD) simulations have shown that the radiation pressure barrier in high-mass star formation can be circumvented if the accretion flow is anisotropic. These studies have shown that there are several ways to supply an anisotropic accretion flow to the high-mass star (see Figure~\ref{fig:hmt}). The first mechanism is by an optically thick axisymmetric accretion disk that funnels gas onto the massive star, allowing radiation to escape along the star’s bipolar directions launching radiation pressure dominated bubbles \citep[e.g., ][]{Yorke2002, Krumholz2009a, Kuiper2013, Rosen2016}. Accretion disk formation is a natural consequence due to the angular momentum content of the collapsing core.  Material is also delivered to the star by self-shielding optically thick filaments that form from the core’s turbulent structure allowing the radiation to escape along low-density channels \citep{Rosen2016}. Additionally, material may also be delivered by radiative Rayleigh-Taylor instabilities that develop at the radiation pressure dominated bubble shells and generate self-shielding Rayleigh-Taylor fingers that can penetrate through these bubbles, potentially onto the star-disk system \citep{Krumholz2009a, Rosen2016, Rosen2019}. We note that this last mechanism for accretion is still heavily debated in the literature \citep{Kuiper2012, Klassen2016, Mignon-Risse2020}. \citet{Rosen2016} argue that this disagreement is related to properly resolving the dense, expanding shells since radiative Rayleigh-Taylor instabilities should arise from small scale perturbations in the dense shells that surround the high-mass protostar \citep{Jacquet2011}.

\subsubsection{Protostellar Outflows}

Similar to low-mass star formation, outflows are ubiquitous in high-mass star formation. Simulations of forming high-mass stars suggest that the momentum injected by outflows entrain and expel 20-50\% of the gas in dense cores \citep{Kuiper2016}. The efficiency of entrainment depends on the mass-loss rate (i.e., the fraction of accreted material that is launched into the outflow) and the core surface density, with lower densities and higher mass-loss rates producing lower star-formation efficiencies \citep{Cunningham2011, Kuiper2016}.

The presence of outflows makes radiative feedback less effective because they evacuate polar cavities of reduced optical depth through the ambient core  \citep{Cunningham2011, Kuiper2016}. These cavities enhance the radiative flux in the poleward direction thereby diminishing the radiative heating and outward radiation force exerted on the protostellar disk and infalling core material in the equatorial direction. Additionally, outflows, in conjunction with photoionization, broadens the outflow cavities that surround accreting massive protostars \citep{Kuiper2018}. If the mass reservoir is finite, as in the TC model, then protostellar outflows alone limit the accretion scenario. However, outflows are likely not powerful enough to shut off the accretion flow onto massive stars if there exists a large mass reservoir around the accreting core \citep{Kuiper2018}. 

\subsubsection{\add{Photoionization}}
Accreting high-mass protostars also emit photoionizing radiation ($E_{\nu}\ge 13.6$ eV) due to the high surface temperatures they attain as they contract to the main sequence. \citet{Tanaka2016} found that photoionization for high-mass star formation becomes important when the star reaches 10-20 \msun and that the stellar mass that this occurs at depends on the accretion rate, which depends on the initial core's surface density, since protostellar evolution (i.e., stellar radius and luminosity) are intimately tied to accretion \citep{Hosokawa2010}. The accretion disk partially self-shields against photoionization and instead the magnetically launched outflows (discussed next) will become photoionized leading to a very small, jet-like UC\hii\ region that is confined by the protostellar outflows, but eventually broadens the bipolar outflow cavities when the star is sufficiently massive \citep{Tan2003, Tanaka2016, Kuiper2018}. The majority of outflows become photoionized quite rapidly ($t\sim10^3-10^4$ yr) after the initial \hii\ region formation \citep{Tanaka2016}. As the \hii\ region expands, the thermal pressure from the warm ionized $\sim 10^4$ K gas squeezes the accretion disk, thereby increasing the accretion rate onto the star temporarily but does not shut off accretion completely \citep{Kuiper2018}.

\subsubsection{Stellar Winds}
The feedback associated with stellar winds has likely received the least theoretical attention in high-mass star formation. Once the high-mass protostar reaches a high surface temperature of $\sim 12.5$\,kK, as it contracts to the main sequence, its high luminosity will drive fast UV line-driven spherical winds from its surface \citep{Vink2001}. The mass-loss rate ($\dot{M}_{\rm \star,\, wind}$) and wind ejection velocity both increase as the star increases in mass and decreases in size. The stellar winds will collide with infalling material and shock-heat gas to $\sim 10^7$ K that will adiabatically expand and push on the infalling gas \citep{Rosen2014, Geen2020}. Similarly, stellar radiation can also produce line-driven winds in the surface layers of the accretion disk within a few stellar radii of the stellar photosphere, ablating the disk. \citet{Kee2019} find that the disk ablation rate is stronger than  the stellar wind mass-loss rate: $\dot{M}_{\rm abl} = 6.5 \pm 1 \dot{M}_{\rm \star,\, wind}$. Their results suggest that disk ablation by line-driven disk winds is a strong feedback effect for very massive (proto)stars and may set the upper mass-limit of the IMF.

\section{Bridging Simulations and Observations through Synthetic Observations}
\label{sec:synthetic}

Synthetic observations provide a critical means to compare numerical simulations to observations, thereby helping to distinguish between theories, infer underlying physical properties and providing physical  interpretation. A {\it synthetic observation} is defined as a model for the emission that would be observed from a simulation if it were a real astronomical object viewed by a particular telescope.  Synthetic observations should reflect observational limitations including noise, resolution and instrumental effects (e.g., interferometry). Here we briefly summarize different modelling approaches and some results from recent synthetic observation studies.  \add{We refer the reader to the review by \citet{Haworth+2018} for a more detailed review of the use of synthetic observations from the literature used to study star formation.}

\subsection{Modelling Approaches}

Most synthetic observations are produced in post-processing, whereby numerical outputs are used as initial conditions for radiative transfer (RT) calculations \add{to deduce the gas and dust temperature distribution, and therefore the resulting dust and line emission}. A variety of public, multi-dimensional RT codes are available, including {\sc hyperion} (dust RT), {\sc moccassin} (dust RT), {\sc radmc-3d} (dust and line RT), {\sc lime} (line and dust RT) and {\sc polaris} (line and dust RT),  which facilitate the production of synthetic observations \citep{Ercolano+2005, BrinchHogerheijde2010, Robitaille2011, Dullemond+2012, Reissl2016}. Other codes are designed to compute time-dependent chemical reactions given some initial abundances, radiative boundary conditions and gas density distribution, such as {\sc 3d-pdr} (3D gas chemistry/photo-dissociation and LTE RT), {\sc torus-3d-pdr} (3D chemistry, photo-dissociation and photo-ionization), {\sc uclchem} (1D gas-grain chemistry), {\sc nautilus } (1D gas-grain chemistry), \add{and {\sc krome} (chemical network modeling)} \citep{Bisbas+2012,Bisbas+2015, Grassi2014a, Ruaud+2016,Holdship+2017}.  

\add{These codes process numerical outputs to produce realistic mock observations by using RT to compute the realistic dust and gas temperature distributions from the simulaiton inputs, however this method is not able to capture the heating and cooling processes in real time. This limitation can be avoided by processing outputs from simulations that include RT and heating and cooling processes, including those attributed to gas dynamics, which therefore includes \textit{on-the-fly} temperature information. Additionally, the initial abundances of molecules and/or dust most be supplied when producing synthetic observations. Such quantities are usually computed from gas properties using simple assumptions (i.e., the X$_{\rm CO}$ factor or dust-to-gas ratio).} In light of these limitations, there is a small but growing number of codes combine hydrodynamics, simple chemistry and RT, such as {\sc flash} and {\sc torus-3D} \citep{Grassi+2014,Harries+2019}.


\subsection{Results from Synthetic Observation Comparisons}

The molecular tracers through which cores and filaments are observed suffer from optical depth effects (e.g., $^{12}$CO and $^{13}$CO) and limited spatial extent (e.g., N$_2$H$^+$ and NH$_3$), which hide underlying physical properties and potentially bias observational results. However, synthetic observations show that even cores in dynamical environments can exhibit sonic velocity dispersions and small velocity offsets between the core center and core envelope \citep{Ayliffe+2007,Offner+2008}. However, at late times simulations that neglect magnetic fields and feedback also exhibit infall velocities that exceed those typical of protostellar cores, while simulations that include these effects obtain better agreement  \citep{Maureira+2017}.  A careful study of core projection effects, which incorporated radiative transfer and chemistry, found that the true core masses, velocity dispersions and virial parameters may differ by $\sim 40$\% from the values inferred from $^{13}$CO observations \citep{Beaumont+2013}.  Analysis and synthetic observations of filaments in simulated star-forming regions indicate that filaments have a distribution of widths, rather than a fixed size of 0.1\,pc, and are frequently comprised of coherent sub-filaments that manifest as multiple emission peaks \citep{Smith+2014,Smith+2016}. 

Synthetic observations also play an important role in revealing the mechanisms of binary star system formation. Catching binary formation in action is a major observational challenge given the small scales and high column densities involved. Synthetic observations 
help constrain the impact of detection limits on attempts to observe forming binaries, which in turn inform the frequency and rate of the two main binary formation channels. 
Synthetic observations indicate that ALMA has sufficient sensitivity and resolution to detect starless cores undergoing turbulent fragmentation, which would otherwise be missed by CARMA and JCMT \citep{Offner+2012,Mairs+2014}. The lack of detected substructure in starless cores then indicates either that these  cores are not yet collapsing or that core fragmentation is rare \citep{Dunham+2016}. However, ALMA and VLA detections of substructure in several cores in Ophiuchus and Perseus are in good agreement with rates derived from synthetic observations of magnetized, turbulent core collapse \citep{Pineda+2015,Kirk+2017a}. Larger observational samples of cores across more star-forming regions are required to place tighter constraints on binary formation via turbulent fragmentation.

On smaller scales, synthetic observations show that disk fragmentation, which produces compact fragments within 10s to 100s of au of the central protostar, as well as spiral structure should also be readily detectable with ALMA \citep{Vorobyov+2013b,Evans+2019, Harries2017, Meyer2017, Ahmadi2019a}. To date a variety of disk spiral structure has been detected but only a few cases of fragmentation has been observed \citep{Tobin+2016,Takami+2018,Ilee2018b,Tobin2018b, Maud2019a, Zapata2019}. Given the challenges of measuring disk masses and radii, synthetic observations are crucial to constrain disk models and derive the underlying disk properties \citep{Dunham+2014c, Ahmadi2019a}.

Synthetic observations play an important role in assessing the accuracy of stellar feedback measurements. Protostellar outflow motions are difficult to disentangle from the background turbulence and thus prone to large uncertainties. However, synthetic observations show that methods for estimating the optical depth by combining multiple tracers and subtracting off the dense core component produce fairly accurate estimates \citep{Offner+2011}. Without optical depth corrections, even high-resolution interferometric observations significantly over-estimate outflow mass and momentum \citep{Bradshaw+2015}. Synthetic observations show the velocity dispersion associated with the outflow is also strongly correlated with the choice of tracer and beam resolution \citep{OffnerArce2014}. In single-dish maps much of the mass and momentum over-estimation of feedback is caused by projection effects whereby foreground and background gas is mapped into the velocity channels that are visually associated with feedback \citep{Xu+2020}. Machine learning approaches trained on dust or line synthetic observations can be used to accurately identify feedback and disentangle emission actually produced by feedback from emission due to line-of-sight contamination \citep{Xu+2017,VanOort+2019,Xu+2020}. 

Given the complex interplay of different feedback mechanisms in high-mass star formation, synthetic observations have played an integral role in determining how high-mass stellar feedback from outflows, photoionization, radiation pressure, and stellar winds shape the ISM from high-mass protostellar cores to massive star clusters. Such synthetic observations have predicted that high mass disks and their internal structure, including fragmentation and spiral structure, should be detectable by ALMA for highly embedded high-mass protostars located at large distances (d\gtrsim 1 kpc) \citep{Krumholz2007a, Harries2017, Meyer2017, Ahmadi2019a}. 
Additionally, synthetic observations demonstrated that the morphologies and evolution of expanding \hii\ regions depends on the interaction of the ionized gas flow with the accretion flows of the surrounding neutral gas \citep{Peters2011}. Quenching of the \hii\ regions by neutral gas causes them to flicker, potentially on observable time scales \citep{Peters2010a, Peters2010b, Peters2011}. On larger scales, synthetic observations allow a direct comparison with observations to determine how feedback from young massive stars shapes the pillars and trunks commonly observed in high-mass star forming regions and evacuate gas from star clusters \citep[e.g.,][]{McLeod2015}.



\section{Summary and Future Outlook}
\label{sec:con}
In this review, we highlighted many of the recent advances in our understanding of star formation through the use of observational and theoretical methods starting from the large-scale complex and filamentary structures of GMCs ($\sim 10s$ pc scales) down to the ($\sim 10s$ au scales) of circumstellar disks. While there are many similarities between low- and high-mass star formation there are also many differences. Low- and high-mass stars form from the gravitational collapse of dense cores that predominately exist in dense filamentary structures, and their feedback (e.g., protostellar outflows and radiative heating) affects the accretion and fragmentation processes of their host cores. However, high-mass stars form predominately in clustered environments whereas low-mass star formation is more isolated. This
difference is likely characterized by the higher densities, velocity dispersions, and dynamical accretion flows from large to small scales inherent to high-mass star-forming regions. Additionally, high-mass star formation is more chaotic than low-mass star formation because the feedback associated with their intense luminosities, outflows, and stellar winds affects their formation and \add{the} dynamics of their natal clouds.

Advances in interferometry and the arrival of ALMA, in particular, have helped to revolutionize the study of low- and high- mass star formation over the last decade. These facilities have enabled the first observations of detailed chemistry and properties of protostellar disks, molecular outflow features and sub-structure in dense cores and clouds beyond a kpc. In the next decade, large-scale surveys by ALMA will significantly increase our understanding of star formation across different environments\add{, including the dynamical importance of magnetic fields across all size scales,} and enhance the statistics of forming high-mass stars. In the far future, observations using the proposed next-generation Very Large Array (ngVLA) and expanded ALMA facilities will push studies of star formation to higher sensitivity and resolution, resolving ever smaller scales and more distant regions. 

Meanwhile, computational method advances, faster supercomputing facilites and new system architectures will enable calculations that follow the full dynamic range -- kpc to au scales -- and multi-physics nature of star formation. Future calculations will be able to begin with realistic galactic-scale initial conditions and follow the gas evolution through the formation of individual stars. Such work is essential to understand in detail how star formation depends on the larger environment, drives the life cycle of galaxies and varies across the broad range of physical conditions observed in the Universe.

Given the complexity of star formation, the use of synthetic observations has bridged the gap between numerical simulations and observations, allowing us to test theories and identify the underlying physical processes that shape the star formation process. Star formation observations depend on the distributions of dust and molecules in the ISM. Therefore, it is crucial to model dust and molecular abundances, including advection, creation and destruction processes, for a variety of molecular species and dust grain size distributions in star formation simulations for direct comparison to observations. Given this importance, star formation simulations are becoming more complex: including radiative transfer, chemical networks of important molecules, and dust evolution. This ongoing and future work will bring greater clarity to current debates about the relationship between low- and high-mass star formation, the role of various physical processes such as magnetic fields and the connection between individual star formation and the stellar initial mass function.

\begin{acknowledgements}
 A.L.R., S.S.R.O, A.B., and E.V.-S. thank the staff of ISSI for their generous hospitality and creating the fruitful cooperation. The authors thank the anonymous referee for their advice and suggestions that greatly improved the manuscript. A.L.R acknowledges support from NASA through Einstein Postdoctoral Fellowship grant No. PF7-180166 awarded by the Chandra X-ray Center, which is operated by the Smithsonian Astrophysical Observatory for NASA under contract NAS8- 03060. S.S.R.O. acknowledges funding from NSF Career grant AST-1650486. E.V.-S. acknowledges financial support from CONACYT grant 255295, Mexico. A.L.R. would like to thank Stefanie Walch and Rolf Kuiper for insightful conversations regarding this review.
\end{acknowledgements}

\bibliographystyle{apj_author.bst}
\bibliography{highmass-lowmass2.bib}

\end{document}